\documentclass{article}

\PassOptionsToPackage{numbers, sort&compress}{natbib}

\usepackage[final]{neurips_2019}

\usepackage{amsmath,amsfonts,bm}

\def\eqref#1{equation~\ref{#1}}

\def\1{\bm{1}}

\def\vv{{\bm{v}}}

\def\vx{{\bm{x}}}
\def\vy{{\bm{y}}}

\def\mA{{\bm{A}}}

\def\mD{{\bm{D}}}
\def\mE{{\bm{E}}}

\def\mG{{\bm{G}}}

\def\mI{{\bm{I}}}

\def\mL{{\bm{L}}}

\def\mS{{\bm{S}}}
\def\mT{{\bm{T}}}
\def\mU{{\bm{U}}}

\def\mX{{\bm{X}}}

\def\mLambda{{\bm{\Lambda}}}

\DeclareMathAlphabet{\mathsfit}{\encodingdefault}{\sfdefault}{m}{sl}
\SetMathAlphabet{\mathsfit}{bold}{\encodingdefault}{\sfdefault}{bx}{n}

\def\emD{{D}}

\usepackage[utf8]{inputenc} 
\usepackage[T1]{fontenc}    
\usepackage{hyperref}       
\usepackage{url}            
\usepackage{booktabs}       
\usepackage{amsfonts}       
\usepackage{nicefrac}       
\usepackage{microtype}      
\usepackage{xcolor}         
\usepackage{mathtools}      
\usepackage{tikz}           
\usepackage{pgf}            
\usepackage{subcaption}     
\usepackage{enumitem}       
\usepackage{soul}           
\usepackage[separate-uncertainty=true,detect-weight=true]{siunitx}  
\usepackage{multirow}       
\usepackage{placeins}       
\usepackage{wrapfig}        

\let\pgfimageWithoutPath\pgfimage
\renewcommand{\pgfimage}[2][]{\pgfimageWithoutPath[#1]{figures/#2}}  

\usetikzlibrary{calc}

\definecolor{uglyblue}{named}{blue}
\definecolor{uglygreen}{named}{green}
\definecolor{uglyred}{named}{red}
\definecolor{uglyyellow}{named}{yellow}

\definecolor{brewerpurple}{RGB}{117,112,179}
\definecolor{brewergreen}{RGB}{27,158,119}
\definecolor{brewerred}{RGB}{217,95,2}

\definecolor{blue}{RGB}{1, 115, 178}
\definecolor{orange}{RGB}{222, 143, 5}
\definecolor{green}{RGB}{2, 158, 115}
\definecolor{red}{RGB}{213, 94, 0}
\definecolor{pink}{RGB}{204, 120, 188}
\definecolor{yellow}{RGB}{236, 225, 51}

\title{Diffusion Improves Graph Learning}

\author{%
  Johannes Gasteiger, Stefan Weißenberger, Stephan Günnemann\\
  Technical University of Munich\\
  \texttt{\{j.gasteiger,stefan.weissenberger,guennemann\}@in.tum.de}\\
}

\begin{document}

\maketitle

\begin{abstract}
    Graph convolution is the core of most Graph Neural Networks (GNNs) and usually approximated by message passing between direct (one-hop) neighbors. In this work, we remove the restriction of using only the direct neighbors by introducing a powerful, yet spatially localized graph convolution: Graph diffusion convolution (GDC). GDC leverages generalized graph diffusion, examples of which are the heat kernel and personalized PageRank. It alleviates the problem of noisy and often arbitrarily defined edges in real graphs. We show that GDC is closely related to spectral-based models and thus combines the strengths of both spatial (message passing) and spectral methods. We demonstrate that replacing message passing with graph diffusion convolution consistently leads to significant performance improvements across a wide range of models on both supervised and unsupervised tasks and a variety of datasets. Furthermore, GDC is not limited to GNNs but can trivially be combined with any graph-based model or algorithm (e.g. spectral clustering) without requiring any changes to the latter or affecting its computational complexity. Our implementation is available online. \footnote{\url{https://www.daml.in.tum.de/gdc}}
\end{abstract}

\section{Introduction} \label{sec:intro}

When people started using graphs for evaluating chess tournaments in the middle of the 19th century they only considered each player's direct opponents, i.e. their first-hop neighbors. Only later was the analysis extended to recursively consider higher-order relationships via $\mA^2$, $\mA^3$, etc. and finally generalized to consider all exponents at once, using the adjacency matrix's dominant eigenvector \citep{landau_zur_1895,vigna_spectral_2016}.
The field of Graph Neural Networks (GNNs) is currently in a similar state. Graph Convolutional Networks (GCNs) \citep{kipf_semi-supervised_2017}, also referred to as Message Passing Neural Networks (MPNNs) \citep{gilmer_neural_2017} are the prevalent approach in this field but they only pass messages between neighboring nodes in each layer. These messages are then aggregated at each node to form the embedding for the next layer. While MPNNs do leverage higher-order neighborhoods in deeper layers, limiting each layer's messages to one-hop neighbors seems arbitrary. Edges in real graphs are often noisy or defined using an arbitrary threshold \citep{tang_atomistic_2018}, so we can clearly improve upon this approach.

Since MPNNs only use the immediate neigborhod information, they are often referred to as spatial methods. On the other hand, spectral-based models do not just rely on first-hop neighbors and capture more complex graph properties \citep{defferrard_convolutional_2016}. However, while being theoretically more elegant, these methods are routinely outperformed by MPNNs on graph-related tasks \citep{kipf_semi-supervised_2017,velickovic_graph_2018,xu_how_2019} and do not generalize to previously unseen graphs. This shows that message passing is a powerful framework worth extending upon.
To reconcile these two separate approaches and combine their strengths we propose a novel technique of performing message passing inspired by spectral methods: Graph diffusion convolution (GDC). Instead of aggregating information only from the first-hop neighbors, GDC aggregates information from a larger neighborhood. This neighborhood is constructed via a new graph generated by sparsifying a generalized form of graph diffusion. We show how graph diffusion is expressed as an equivalent polynomial filter and how GDC is closely related to spectral-based models while addressing their shortcomings. GDC is spatially localized, scalable, can be combined with message passing, and generalizes to unseen graphs. Furthermore, since GDC generates a new sparse graph it is not limited to MPNNs and can trivially be combined with \emph{any} existing graph-based model or algorithm in a plug-and-play manner, i.e. without requiring changing the model or affecting its computational complexity. We show that GDC consistently improves performance across a wide range of models on both supervised and unsupervised tasks and various homophilic datasets. In summary, this paper's core contributions are:
\setlist{nolistsep}
\begin{enumerate}[leftmargin=*,itemsep=2pt]
    \item Proposing graph diffusion convolution (GDC), a more powerful and general, yet spatially localized alternative to message passing that uses a sparsified generalized form of graph diffusion. GDC is not limited to GNNs and can be combined with any graph-based model or algorithm.
    \item Analyzing the spectral properties of GDC and graph diffusion. We show how graph diffusion is expressed as an equivalent polynomial filter and analyze GDC's effect on the graph spectrum.
    \item Comparing and evaluating several specific variants of GDC and demonstrating its wide applicability to supervised and unsupervised learning on graphs.
\end{enumerate}


\section{Generalized graph diffusion} \label{sec:graphdiff}

We consider an undirected graph $\mathcal{G} = (\mathcal{V}, \mathcal{E})$ with node set $\mathcal{V}$ and edge set $\mathcal{E}$. We denote with $N = |\mathcal{V}|$ the number of nodes and $\mA \in \mathbb{R}^{N \times N}$ the adjacency matrix. We define generalized graph diffusion via the diffusion matrix
\begin{equation}
    \label{eq:graphdiff}
    \mS = \sum_{k=0}^{\infty} \theta_k \mT^k,
\end{equation}
with the weighting coefficients $\theta_k$, and the generalized transition matrix $\mT$. The choice of $\theta_k$ and $\mT^k$ must at least ensure that Eq. \ref{eq:graphdiff} converges. In this work we will consider somewhat stricter conditions and require that $\sum_{k=0}^{\infty} \theta_k = 1$, $\theta_k \in [0, 1]$, and that the eigenvalues of $\mT$ are bounded by $\lambda_i \in [0, 1]$, which together are sufficient to guarantee convergence. Note that regular graph diffusion commonly requires $\mT$ to be column- or row-stochastic.

\textbf{Transition matrix.} Examples for $\mT$ in an undirected graph include the random walk transition matrix $\mT_\text{rw} = \mA \mD^{-1}$ and the symmetric transition matrix $\mT_\text{sym} = \mD^{-1/2} \mA \mD^{-1/2}$, where the degree matrix $\mD$ is the diagonal matrix of node degrees, i.e. $\emD_{ii} = \sum_{j=1}^N A_{ij}$. Note that in our definition $\mT_\text{rw}$ is column-stochastic.
We furthermore adjust the random walk by adding (weighted) self-loops to the original adjacency matrix, i.e. use $\tilde{\mT}_\text{sym} = (w_{\text{loop}} \mI_N + \mD)^{-1/2} (w_{\text{loop}} \mI_N + \mA) (w_{\text{loop}} \mI_N + \mD)^{-1/2}$, with the self-loop weight $w_{\text{loop}} \in \mathbb{R}^+$. This is equivalent to performing a lazy random walk with a probability of staying at node $i$ of $p_{\text{stay}, i} = w_{\text{loop}}/\emD_{i}$.

\textbf{Special cases.} Two popular examples of graph diffusion are personalized PageRank (PPR) \citep{page_pagerank_1998} and the heat kernel \citep{kondor_diffusion_2002}. PPR corresponds to choosing $\mT = \mT_\text{rw}$ and $\theta^\text{PPR}_k = \alpha (1 - \alpha)^k$, with teleport probability $\alpha \in (0, 1)$ \citep{chung_heat_2007}. The heat kernel uses $\mT = \mT_\text{rw}$ and $\theta^\text{HK}_k = e^{-t} \frac{t^k}{k!}$, with the diffusion time $t$ \citep{chung_heat_2007}. Another special case of generalized graph diffusion is the approximated graph convolution introduced by \citet{kipf_semi-supervised_2017}, which translates to $\theta_1 = 1$ and $\theta_k = 0$ for $k \neq 1$ and uses $\mT = \tilde{\mT}_{\text{sym}}$ with $w_{\text{loop}} = 1$.

\textbf{Weighting coefficients.} We compute the series defined by Eq. \ref{eq:graphdiff} either in closed-form, if possible, or by restricting the sum to a finite number $K$. Both the coefficients defined by PPR and the heat kernel give a closed-form solution for this series that we found to perform well for the tasks considered. Note that we are not restricted to using $\mT_\text{rw}$ and can use any generalized transition matrix along with the coefficients $\theta^\text{PPR}_k$ or $\theta^\text{HK}_k$ and the series still converges.
We can furthermore choose $\theta_k$ by repurposing the graph-specific coefficients obtained by methods that optimize coefficients analogous to $\theta_k$ as part of their training process. We investigated this approach using label propagation \citep{chen_adaptive_2013,berberidis_adaptive_2019} and node embedding models \citep{abu-el-haija_watch_2018}. However, we found that the simple coefficients defined by PPR or the heat kernel perform better than those learned by these models (see Fig. \ref{fig:adadif} in Sec. \ref{sec:exp}).

\section{Graph diffusion convolution} \label{sec:gdc}

\begin{figure}
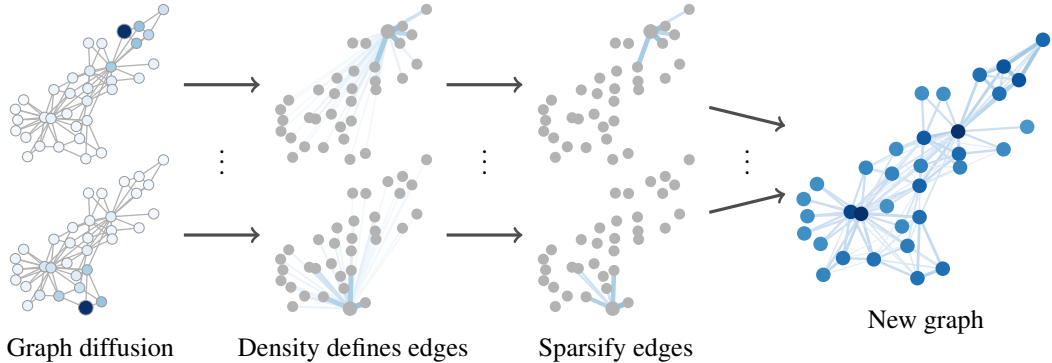

\centering
\begin{tikzpicture}
    \node[anchor=west] (graph1) at (0, 1) {\input{figures/graph_ppr1.pgf}};
    \node[anchor=west] (edges1) at ($0.25*(\textwidth, 0) + (0, 1)$) {\input{figures/edges_ppr1.pgf}};
    \node[anchor=west] (sparse1) at ($0.5*(\textwidth, 0) + (0, 1)$) {\input{figures/edges_ppr_sparse1.pgf}};
    \node[anchor=west] (graph2) at (0, -1) {\input{figures/graph_ppr2.pgf}};
    \node[anchor=west] (edges2) at ($0.25*(\textwidth, 0) - (0, 1)$) {\input{figures/edges_ppr2.pgf}};
    \node[anchor=west] (sparse2) at ($0.5*(\textwidth, 0) - (0, 1)$) {\input{figures/edges_ppr_sparse2.pgf}};
    \node[anchor=west] (gdc) at ($0.75*(\textwidth, 0)$) {\input{figures/graph_gdc.pgf}};

    \draw[->,line width=1.2pt,color=black!70] (graph1) -- node[midway] (ge1) {} (edges1);
    \draw[->,line width=1.2pt,color=black!70] (edges1) -- node[midway] (es1) {} (sparse1);
    \draw[->,line width=1.2pt,color=black!70] (graph2) -- node[midway] (ge2) {} (edges2);
    \draw[->,line width=1.2pt,color=black!70] (edges2) -- node[midway] (es2) {} (sparse2);
    \draw[->,line width=1.2pt,color=black!70] (sparse1) -- node[midway] (sg1) {} (gdc);
    \draw[->,line width=1.2pt,color=black!70] (sparse2) -- node[midway] (sg2) {} (gdc);

    \node[rotate=90] at ($0.5*(ge1) + 0.5*(ge2)$) {\dots};
    \node[rotate=90] at ($0.5*(es1) + 0.5*(es2)$) {\dots};
    \node[rotate=90] at ($0.5*(sg1) + 0.5*(sg2)$) {\dots};


    \node[anchor=north] at (graph2.south) {Graph diffusion};
    \node[anchor=north] at (edges2.south) {Density defines edges};
    \node[anchor=north] at (sparse2.south) {Sparsify edges};
    \node[anchor=north] at (gdc.south) {New graph};
\end{tikzpicture}

\caption{Illustration of graph diffusion convolution (GDC). We transform a graph $\mA$ via graph diffusion and sparsification into a new graph $\tilde{\mS}$ and run the given model on this graph instead.}
\label{fig:gdc}
\end{figure}

Essentially, graph diffusion convolution (GDC) exchanges the normal adjacency matrix $\mA$ with a sparsified version $\tilde{\mS}$ of the generalized graph diffusion matrix $\mS$, as illustrated by Fig. \ref{fig:gdc}. This matrix defines a weighted and directed graph, and the model we aim to augment is applied to this graph instead. We found that the calculated edge weights are beneficial for the tasks considered. However, we even found that GDC works when ignoring the weights after sparsification. This enables us to use GDC with models that only support unweighted edges such as the degree-corrected stochastic block model (DCSBM). If required, we make the graph undirected by using $(\tilde{\mS} + \tilde{\mS}^T)/2$, e.g. for spectral clustering. With these adjustments GDC is applicable to \emph{any} graph-based model or algorithm.

\textbf{Intuition.} The general intuition behind GDC is that graph diffusion smooths out the neighborhood over the graph, acting as a kind of denoising filter similar to Gaussian filters on images. This helps with graph learning since both features and edges in real graphs are often noisy. Previous works also highlighted the effectiveness of graph denoising. \citet{berberidis_node_2018} showed that PPR is able to reconstruct the underlying probability matrix of a sampled stochastic block model (SBM) graph. \citet{kloumann_block_2017} and \citet{ragain_community_2017} showed that PPR is optimal in recovering the SBM and DCSBM clusters in the space of landing probabilities under the mean field assumption. \citet{li_optimizing_2019} generalized this result by analyzing the convergence of landing probabilities to their mean field values. These results confirm the intuition that graph diffusion-based smoothing indeed recovers meaningful neighborhoods from noisy graphs.

\textbf{Sparsification.} Most graph diffusions result in a dense matrix $\mS$. This happens even if we do not sum to $k = \infty$ in Eq. \ref{eq:graphdiff} due to the ``four/six degrees of separation'' in real-world graphs \citep{backstrom_four_2012}. However, the values in $\mS$ represent the influence between all pairs of nodes, which typically are highly localized \citep{nassar_strong_2015}. This is a major advantage over spectral-based models since the spectral domain does not provide any notion of locality. Spatial localization allows us to simply truncate small values of $\mS$ and recover sparsity, resulting in the matrix $\tilde{\mS}$. In this work we consider two options for sparsification: 1. top-$k$: Use the $k$ entries with the highest mass per column, 2. Threshold $\epsilon$: Set entries below $\epsilon$ to zero.
Sparsification would still require calculating a dense matrix $\mS$ during preprocessing. However, many popular graph diffusions can be approximated efficiently and accurately in linear time and space. Most importantly, there are fast approximations for both PPR \citep{andersen_local_2006,wei_topppr:_2018} and the heat kernel \citep{kloster_heat_2014}, with which GDC achieves a linear runtime $\mathcal{O}(N)$.
Furthermore, top-$k$ truncation generates a regular graph, which is amenable to batching methods and solves problems related to widely varying node degrees \citep{decelle_inference_2011}. Empirically, we even found that sparsification slightly \emph{improves} prediction accuracy (see Fig. \ref{fig:degree} in Sec. \ref{sec:exp}). After sparsification we calculate the (symmetric or random walk) transition matrix on the resulting graph via $\mT^{\tilde{\mS}}_\text{sym} = \mD_{\tilde{\mS}}^{-1/2} \tilde{\mS} \mD_{\tilde{\mS}}^{-1/2}$.

\textbf{Limitations.} GDC is based on the assumption of homophily, i.e. ``birds of a feather flock together'' \citep{mcpherson_birds_2001}. Many methods share this assumption and most common datasets adhere to this principle. However, this is an often overlooked limitation and it seems non-straightforward to overcome. One way of extending GDC to heterophily, i.e. ``opposites attract'', might be negative edge weights \citep{ma_diffusion_2016,derr_signed_2018}. Furthermore, we suspect that GDC does not perform well in settings with more complex edges (e.g. knowledge graphs) or graph reconstruction tasks such as link prediction. Preliminary experiments showed that GDC indeed does not improve link prediction performance.

\section{Spectral analysis of GDC} \label{sec:spectral}

Even though GDC is a spatial-based method it can also be interpreted as a graph convolution and analyzed in the graph spectral domain. In this section we show how generalized graph diffusion is expressed as an equivalent polynomial filter and vice versa. Additionally, we perform a spectral analysis of GDC, which highlights the tight connection between GDC and spectral-based models.

\textbf{Spectral graph theory.} To employ the tools of spectral theory to graphs we exchange the regular Laplace operator with either the unnormalized Laplacian $\mL_{\text{un}} = \mD - \mA$, the random-walk normalized $\mL_{\text{rw}} = \mI_N - \mT_\text{rw}$, or the symmetric normalized graph Laplacian $\mL_{\text{sym}} = \mI_N - \mT_\text{sym}$ \citep{von_luxburg_tutorial_2007}. The Laplacian's eigendecomposition is $\mL = \mU \mLambda \mU^T$, where both $\mU$ and $\mLambda$ are real-valued. The graph Fourier transform of a vector $\vx$ is then defined via $\hat{\vx} = \mU^T \vx$ and its inverse as $\vx = \mU \hat{\vx}$. Using this we define a graph convolution on $\mathcal{G}$ as $\vx *_{\mathcal{G}} \vy = \mU((\mU^T \vx) \odot (\mU^T \vy))$, where $\odot$ denotes the Hadamard product. Hence, a filter $g_\xi$ with parameters $\xi$ acts on $\vx$ as $g_\xi(\mL) \vx = \mU \hat{\mG}_\xi(\mLambda) \mU^T \vx$, where $\hat{\mG}_\xi(\mLambda) = \text{diag}(\hat{g}_{\xi,1}(\mLambda), \ldots, \hat{g}_{\xi,N}(\mLambda))$. A common choice for $g_\xi$ in the literature is a polynomial filter of order $J$, since it is localized and has a limited number of parameters \citep{hammond_wavelets_2011,defferrard_convolutional_2016}:
\begin{equation}
    g_\xi(\mL) = \sum_{j=0}^J \xi_j \mL^j = \mU \left( \sum_{j=0}^J \xi_j \mLambda^j \right) \mU^T.
    \label{eq:filter}
\end{equation}
\textbf{Graph diffusion as a polynomial filter.} Comparing Eq. \ref{eq:graphdiff} with Eq. \ref{eq:filter} shows the close relationship between polynomial filters and generalized graph diffusion since we only need to exchange $\mL$ by $\mT$ to go from one to the other. To make this relationship more specific and find a direct correspondence between GDC with $\theta_k$ and a polynomial filter with parameters $\xi_j$ we need to find parameters that solve
\begin{equation}
    \sum_{j=0}^{J} \xi_j \mL^j \overset{!}{=} \sum_{k=0}^{K} \theta_k \mT^k.
    \label{eq:spectralandgdc}
\end{equation}
To find these parameters we choose the Laplacian corresponding to $\mL = \mI_n - \mT$, resulting in (see App. \ref{app:difffilter})
\begin{equation}
    \xi_j = \sum_{k=j}^{K} \binom{k}{j} (-1)^j \theta_k, \hspace{1.5cm} \theta_k = \sum_{j=k}^{J} \binom{j}{k} (-1)^k \xi_j, \label{eq:gdctospectral}
\end{equation}
which shows the direct correspondence between graph diffusion and spectral methods. Note that we need to set $J = K$. Solving Eq. \ref{eq:gdctospectral} for the coefficients corresponding to the heat kernel $\theta_k^\text{HK}$ and PPR $\theta_k^\text{PPR}$ leads to
\begin{equation}
    \xi^{\text{HK}}_j = \frac{(-t)^j}{j!},\hspace{1cm} \xi^{\text{PPR}}_j = \left( 1 - \frac{1}{\alpha} \right)^j,
    \label{eq:hk_ppr_spectral}
\end{equation}
showing how the heat kernel and PPR are expressed as polynomial filters. Note that PPR's corresponding polynomial filter converges only if $\alpha > 0.5$. This is caused by changing the order of summation when deriving $\xi^{\text{PPR}}_j$, which results in an alternating series. However, if the series does converge it gives the exact same transformation as the equivalent graph diffusion.

\textbf{Spectral properties of GDC.} We will now extend the discussion to all parts of GDC and analyze how they transform the graph Laplacian's eigenvalues. GDC consists of four steps: 1. Calculate the transition matrix $\mT$, 2. take the sum in Eq. \ref{eq:graphdiff} to obtain $\mS$, 3. sparsify the resulting matrix by truncating small values, resulting in $\tilde{\mS}$, and 4. calculate the transition matrix $\mT_{\tilde{\mS}}$.

\textbf{1. Transition matrix.} Calculating the transition matrix $\mT$ only changes which Laplacian matrix we use for analyzing the graph's spectrum, i.e. we use $\mL_{\text{sym}}$ or $\mL_{\text{rw}}$ instead of $\mL_{\text{un}}$. Adding self-loops to obtain $\tilde{\mT}$ does not preserve the eigenvectors and its effect therefore cannot be calculated precisely. \citet{wu_simplifying_2019} empirically found that adding self-loops shrinks the graph's eigenvalues.

\begin{figure}
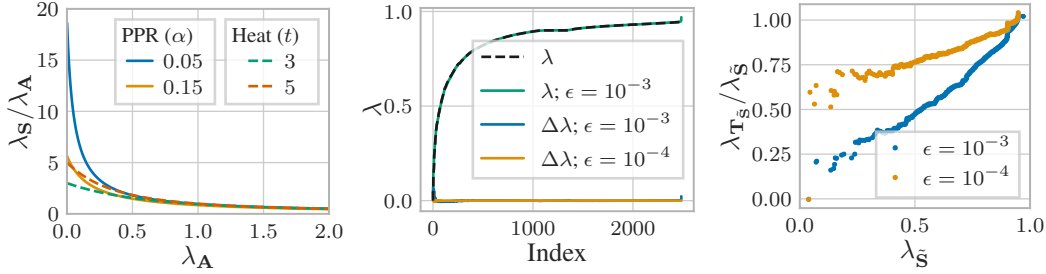

    \centering
    \begin{subfigure}[t]{0.325\textwidth}
        \input{figures/diffusion_filter_amp.pgf}
        \caption{Graph diffusion as a filter, PPR with $\alpha$ and heat kernel with $t$. Both act as low-pass filters.}
        \label{fig:filter:diff}
    \end{subfigure}
    \hfill
    \begin{subfigure}[t]{0.325\textwidth}
        \input{figures/sparsification.pgf}
        \caption{Sparsification with threshold $\epsilon$ of PPR ($\alpha = 0.1$) on \textsc{Cora}. Eigenvalues are almost unchanged.}
        \label{fig:filter:sparsif}
    \end{subfigure}
    \hfill
    \begin{subfigure}[t]{0.325\textwidth}
        \input{figures/transition2_amp.pgf}
        \caption{Transition matrix on \textsc{Cora}'s sparsified graph $\tilde{\mS}$. This acts as a weak high-pass filter.}
        \label{fig:filter:transition}
    \end{subfigure}
    \caption{Influence of different parts of GDC on the Laplacian's eigenvalues $\lambda$.}
    \label{fig:filter}
\end{figure}

\textbf{2. Sum over $\mT^k$.} Summation does not affect the eigenvectors of the original matrix, since $\mT^k \vv_i = \lambda_i \mT^{k-1} \vv_i = \lambda_i^k \vv_i$, for the eigenvector $\vv_i$ of $\mT$ with associated eigenvalue $\lambda_i$. This also shows that the eigenvalues are transformed as
\begin{equation}
    \tilde{\lambda}_i = \sum_{k=0}^{\infty} \theta_k \lambda_i^k.
\end{equation}
Since the eigenvalues of $\mT$ are bounded by 1 we can use the geometric series to derive a closed-form expression for PPR, i.e. $\tilde{\lambda}_i = \alpha \sum_{k=0}^{\infty} (1 - \alpha)^k \lambda_i^k = \frac{\alpha}{1 - (1 - \alpha) \lambda_i}$. For the heat kernel we use the exponential series, resulting in $\tilde{\lambda}_i = e^{-t} \sum_{k=0}^{\infty} \frac{t^k}{k!} \lambda_i^k = e^{t(\lambda_i - 1)}$. How this transformation affects the corresponding Laplacian's eigenvalues is illustrated in Fig. \ref{fig:filter:diff}. Both PPR and the heat kernel act as low-pass filters. Low eigenvalues corresponding to large-scale structure in the graph (e.g.\ clusters \citep{ng_spectral_2002}) are amplified, while high eigenvalues corresponding to fine details but also noise are suppressed.

\textbf{3. Sparsification.} Sparsification changes both the eigenvalues and the eigenvectors, which means that there is no direct correspondence between the eigenvalues of $\mS$ and $\tilde{\mS}$ and we cannot analyze its effect analytically. However, we can use eigenvalue perturbation theory (\citet{stewart_matrix_1990}, Corollary 4.13) to derive the upper bound
\begin{equation}
    \sqrt{\sum_{i=1}^N (\tilde{\lambda}_i - \lambda_i)^2} \leq ||\mE||_\text{F} \leq N ||\mE||_{\max} \leq N \epsilon,
\end{equation}
with the perturbation matrix $\mE = \tilde{\mS} - \mS$ and the threshold $\epsilon$. This bound significantly overestimates the perturbation since PPR and the heat kernel both exhibit strong localization on real-world graphs and hence the change in eigenvalues empirically does not scale with $N$ (or, rather, $\sqrt{N}$). By ordering the eigenvalues we see that, empirically, the typical thresholds for sparsification have almost no effect on the eigenvalues, as shown in Fig. \ref{fig:filter:sparsif} and in the close-up in Fig. \ref{fig:app:sparsif_diff} in App. \ref{app:results}. We find that the small changes caused by sparsification mostly affect the highest and lowest eigenvalues. The former correspond to very large clusters and long-range interactions, which are undesired for local graph smoothing. The latter correspond to spurious oscillations, which are not helpful for graph learning either and most likely affected because of the abrupt cutoff at $\epsilon$.

\textbf{4. Transition matrix on $\tilde{\mS}$.} As a final step we calculate the transition matrix on the resulting graph $\tilde{\mS}$. This step does not just change which Laplacian we consider since we have already switched to using the transition matrix in step 1. It furthermore does not preserve the eigenvectors and is thus again best investigated empirically by ordering the eigenvalues. Fig. \ref{fig:filter:transition} shows that, empirically, this step slightly dampens low eigenvalues. This may seem counterproductive. However, the main purpose of using the transition matrix is ensuring that sparsification does not cause nodes to be treated differently by losing a different number of adjacent edges. The filtering is only a side-effect.

\textbf{Limitations of spectral-based models.} While there are tight connections between GDC and spectral-based models, GDC is actually spatial-based and therefore does not share their limitations. Similar to polynomial filters, GDC does not compute an expensive eigenvalue decomposition, preserves locality on the graph and is not limited to a single graph after training, i.e. typically the same coefficients $\theta_k$ can be used across graphs. The choice of coefficients $\theta_k$ depends on the type of graph at hand and does not change significantly between similar graphs. Moreover, the hyperparameters $\alpha$ of PPR and $t$ of the heat kernel usually fall within a narrow range that is rather insensitive to both the graph and model (see Fig. \ref{fig:hyppar_coeff} in Sec. \ref{sec:exp}).

\section{Related work} \label{sec:related}

Graph diffusion and random walks have been extensively studied in classical graph learning \citep{kondor_diffusion_2002,lafon_diffusion_2006,chen_adaptive_2013,chung_heat_2007}, especially for clustering \citep{kloster_heat_2014}, semi-supervised classification \citep{fouss_experimental_2012,buchnik_bootstrapped_2018}, and recommendation systems \citep{ma_diffusion_2016}. For an overview of existing methods see \citet{masuda_random_2017} and \citet{fouss_experimental_2012}.

The first models similar in structure to current Graph Neural Networks (GNNs) were proposed by \citet{sperduti_supervised_1997} and \citet{baskin_neural_1997}, and the name GNN first appeared in \citep{gori_new_2005,scarselli_graph_2009}. However, they only became widely adopted in recent years, when they started to outperform classical models in many graph-related tasks \citep{duvenaud_convolutional_2015,gasteiger_predict_2019,ying_graph_2018,li_diffusion_2018}. In general, GNNs are classified into spectral-based models \citep{defferrard_convolutional_2016,kipf_semi-supervised_2017,bruna_spectral_2014,henaff_deep_2015,li_adaptive_2018}, which are based on the eigendecomposition of the graph Laplacian, and spatial-based methods \citep{gilmer_neural_2017,hamilton_inductive_2017,li_gated_2016,velickovic_graph_2018,monti_geometric_2017,niepert_learning_2016,pham_column_2017}, which use the graph directly and form new representations by aggregating the representations of a node and its neighbors. However, this distinction is often rather blurry and many models can not be clearly attributed to one type or the other.
Deep learning also inspired a variety of unsupervised node embedding methods. Most models use random walks to learn node embeddings in a similar fashion as word2vec \citep{mikolov_distributed_2013} \citep{perozzi_deepwalk:_2014,grover_node2vec:_2016} and have been shown to implicitly perform a matrix factorization \citep{qiu_network_2018}. Other unsupervised models learn Gaussian distributions instead of vectors \citep{bojchevski_deep_2018}, use an auto-encoder \citep{kipf_variational_2016}, or train an encoder by maximizing the mutual information between local and global embeddings \citep{velickovic_deep_2019}.

There have been some isolated efforts of using extended neighborhoods for aggregation in GNNs and graph diffusion for node embeddings. PPNP \citep{gasteiger_predict_2019} propagates the node predictions generated by a neural network using personalized PageRank, DCNN \citep{atwood_diffusion-convolutional_2016} extends node features by concatenating features aggregated using the transition matrices of $k$-hop random walks, GraphHeat \citep{xu_graph_2019} uses the heat kernel and PAN \citep{ma_pan:_2019} the transition matrix of maximal entropy random walks to aggregate over nodes in each layer, PinSage \citep{ying_graph_2018} uses random walks for neighborhood aggregation, and MixHop \citep{abu-el-haija_mixhop:_2019} concatenates embeddings aggregated using the transition matrices of $k$-hop random walks before each layer. VERSE \citep{tsitsulin_verse:_2018} learns node embeddings by minimizing KL-divergence from the PPR matrix to a low-rank approximation. Attention walk \citep{abu-el-haija_watch_2018} uses a similar loss to jointly optimize the node embeddings and diffusion coefficients $\theta_k$. None of these works considered sparsification, generalized graph diffusion, spectral properties, or using preprocessing to generalize across models.

\section{Experimental results} \label{sec:exp}

\textbf{Experimental setup.} We take extensive measures to prevent any kind of bias in our results. We optimize the hyperparameters of \emph{all} models on \emph{all} datasets with both the unmodified graph and all GDC variants \emph{separately} using a combination of grid and random search on the validation set. Each result is averaged across 100 data splits and random initializations for supervised tasks and 20 random initializations for unsupervised tasks, as suggested by \citet{gasteiger_predict_2019} and \citet{shchur_pitfalls_2018}. We report performance on a test set that was used exactly \emph{once}. We report all results as averages with \SI{95}{\percent} confidence intervals calculated via bootstrapping.

We use the symmetric transition matrix with self-loops $\tilde{\mT}_\text{sym} = (\mI_N + \mD)^{-1/2} (\mI_N + \mA) (\mI_N + \mD)^{-1/2}$ for GDC and the column-stochastic transition matrix $\mT^{\tilde{\mS}}_\text{rw} = \tilde{\mS} \mD_{\tilde{\mS}}^{-1}$ on $\tilde{\mS}$. We present two simple and effective choices for the coefficients $\theta_k$: The heat kernel and PPR. The diffusion matrix $\mS$ is sparsified using either an $\epsilon$-threshold or top-$k$.

\textbf{Datasets and models.} We evaluate GDC on six datasets: The citation graphs \textsc{Citeseer} \citep{sen_collective_2008}, \textsc{Cora} \citep{mccallum_automating_2000}, and \textsc{PubMed} \citep{namata_query-driven_2012}, the co-author graph \textsc{Coauthor CS} \citep{shchur_pitfalls_2018}, and the co-purchase graphs \textsc{Amazon Computers} and \textsc{Amazon Photo} \citep{mcauley_image-based_2015,shchur_pitfalls_2018}. We only use their largest connected components.
We show how GDC impacts the performance of 9 models: Graph Convolutional Network (GCN) \citep{kipf_semi-supervised_2017}, Graph Attention Network (GAT) \citep{velickovic_graph_2018}, jumping knowledge network (JK) \citep{xu_representation_2018}, Graph Isomorphism Network (GIN) \citep{xu_how_2019}, and ARMA \citep{bianchi_graph_2019} are supervised models. The degree-corrected stochastic block model (DCSBM) \citep{karrer_stochastic_2011}, spectral clustering (using $\mL_\text{sym}$) \citep{ng_spectral_2002}, DeepWalk \citep{perozzi_deepwalk:_2014}, and Deep Graph Infomax (DGI) \citep{velickovic_deep_2019} are unsupervised models. Note that DGI uses node features while other unsupervised models do not. We use $k$-means clustering to generate clusters from node embeddings. Dataset statistics and hyperparameters are reported in App. \ref{app:exp}.

\begin{figure}
    \centering
    \input{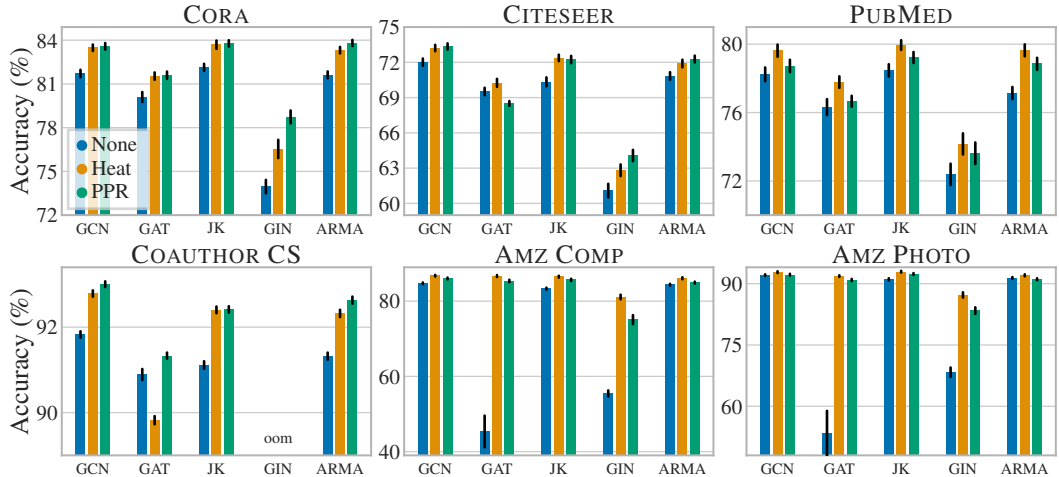}
    \caption{Node classification accuracy of GNNs with and without GDC. GDC consistently improves accuracy across models and datasets. It is able to fix models whose accuracy otherwise breaks down.}
    \label{fig:semi_sup}
\end{figure}

\textbf{Semi-supervised node classification.} In this task the goal is to label nodes based on the graph, node features $\mX \in \mathbb{R}^{N \times F}$ and a subset of labeled nodes $\vy$. The main goal of GDC is improving the performance of MPNN models. Fig. \ref{fig:semi_sup} shows that GDC consistently and significantly improves the accuracy of a wide variety of state-of-the-art models across multiple diverse datasets. Note how GDC is able to fix the performance of GNNs that otherwise break down on some datasets (e.g. GAT). We also surpass or match the previous state of the art on all datasets investigated (see App. \ref{app:results}).

\begin{figure}
    \centering
    \input{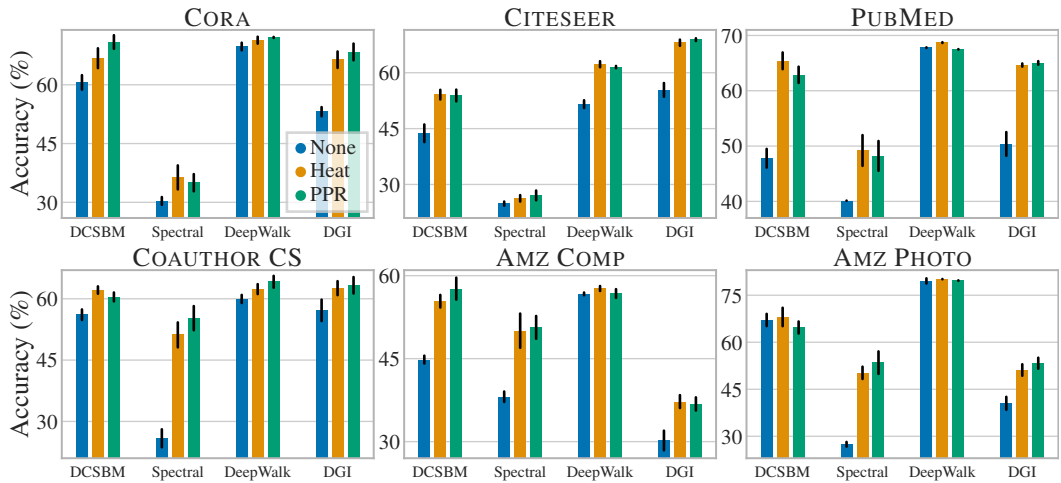}
    \caption{Clustering accuracy with and without GDC. GDC consistently improves the accuracy across a diverse set of models and datasets.}
    \label{fig:clustering}
\end{figure}

\textbf{Clustering.} We highlight GDC's ability to be combined with any graph-based model by reporting the performance of a diverse set of models that use a wide range of paradigms. Fig. \ref{fig:clustering} shows the unsupervised accuracy obtained by matching clusters to ground-truth classes using the Hungarian algorithm. Accuracy consistently and significantly improves for all models and datasets. Note that spectral clustering uses the graph's eigenvectors, which are not affected by the diffusion step itself. Still, its performance improves by up to \num{30} percentage points. Results in tabular form are presented in App. \ref{app:results}.

In this work we concentrate on node-level prediction tasks in a transductive setting. However, GDC can just as easily be applied to inductive problems or different tasks like graph classification. In our experiments we found promising, yet not as consistent results for graph classification (e.g. \num{+2.5} percentage points with GCN on the DD dataset \citep{dobson_distinguishing_2003}). We found no improvement for the inductive setting on PPI \citep{menche_uncovering_2015}, which is rather unsurprising since the underlying data used for graph construction already includes graph diffusion-like mechanisms (e.g. regulatory interactions, protein complexes, and metabolic enzyme-coupled interactions).
We furthermore conducted experiments to answer five important questions:

\begin{figure}
    \centering
    \begin{minipage}[t]{0.325\textwidth}
        \input{figures/sw_degree_accuracy.pgf}
        \caption{GCN+GDC accuracy (using PPR and top-$k$). Lines indicate original accuracy and degree. GDC surpasses original accuracy at around the same degree independent of dataset. Sparsification often improves accuracy.}
        \label{fig:degree}
    \end{minipage}
    \hfill
    \begin{minipage}[t]{0.325\textwidth}
        \input{figures/sw_self_loops.pgf}
        \caption{Difference in GCN+GDC accuracy (using PPR and top-$k$, percentage points) compared to the symmetric $\mT_\text{sym}$ without self-loops. $\mT_\text{rw}$ performs worse and self-loops have no significant effect.}
        \label{fig:transmatrix}
    \end{minipage}
    \hfill
    \begin{minipage}[t]{0.325\textwidth}
        \input{figures/sw_adadif.pgf}
        \caption{Accuracy of GDC with coefficients $\theta_k$ defined by PPR and learned by AdaDIF. Simple PPR coefficients consistently perform better than those obtained by AdaDIF, even with optimized regularization.}
        \label{fig:adadif}
    \end{minipage}
\end{figure}

\textbf{Does GDC increase graph density?} When sparsifying the generalized graph diffusion matrix $\mS$ we are free to choose the resulting level of sparsity in $\tilde{\mS}$. Fig. \ref{fig:degree} indicates that, surprisingly, GDC requires roughly the same average degree to surpass the performance of the original graph independent of the dataset and its average degree ($\epsilon$-threshold in App. \ref{app:results}, Fig. \ref{fig:app:degreeeps}). This suggests that the sparsification hyperparameter can be obtained from a fixed average degree. Note that \textsc{Cora} and \textsc{Citeseer} are both small graphs with low average degree. Graphs become denser with size \citep{leskovec_graphs_2005} and in practice we expect GDC to typically \emph{reduce} the average degree at constant accuracy. Fig. \ref{fig:degree} furthermore shows that there is an optimal degree of sparsity above which the accuracy decreases. This indicates that sparsification is not only computationally beneficial but also improves prediction performance.

\textbf{How to choose the transition matrix $\mT$?} We found $\mT_\text{sym}$ to perform best across datasets. More specifically, Fig. \ref{fig:transmatrix} shows that the symmetric version on average outperforms the random walk transition matrix $\mT_\text{rw}$. This figure also shows that GCN accuracy is largely insensitive to self-loops when using $\mT_\text{sym}$ -- all changes lie within the estimated uncertainty. However, we did find that other models, e.g. GAT, perform better with self-loops (not shown).

\begin{figure}
    \centering
    \input{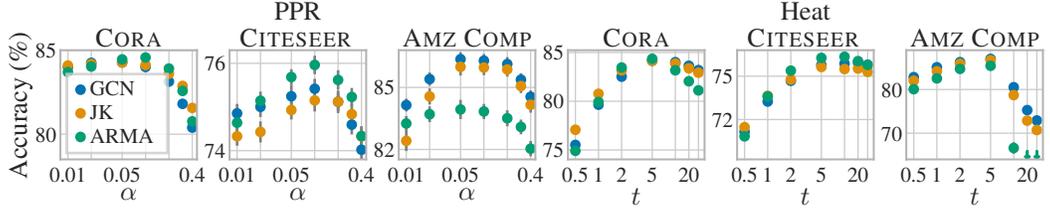}
    \caption{Accuracy achieved by using GDC with varying hyperparameters of PPR ($\alpha$) and the heat kernel ($t$). Optimal values fall within a narrow range that is consistent across datasets and models.}
    \label{fig:hyppar_coeff}
\end{figure}

\textbf{How to choose the coefficients $\theta_k$?} We found the coefficients defined by PPR and the heat kernel to be effective choices for $\theta_k$. Fig. \ref{fig:hyppar_coeff} shows that their optimal hyperparameters typically fall within a narrow range of $\alpha \in [0.05, 0.2]$ and $t \in [1, 10]$. We also tried obtaining $\theta_k$ from models that learn analogous coefficients \citep{chen_adaptive_2013,berberidis_adaptive_2019,abu-el-haija_watch_2018}. However, we found that $\theta_k$ obtained by these models tend to converge to a minimal neighborhood, i.e. they converge to $\theta_0 \approx 1$ or $\theta_1 \approx 1$ and all other $\theta_k \approx 0$.
\begin{wrapfigure}[15]{r}{0.325\textwidth}
    \input{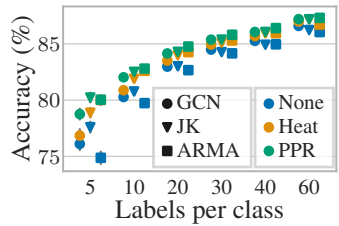}
    \caption{Accuracy on Cora with different label rates. Improvement from GDC increases for sparser label rates.}
    \label{fig:label_rate}
\end{wrapfigure}
This is caused by their training losses almost always decreasing when the considered neighborhood shrinks.
We were able to control this overfitting to some degree using strong regularization (specifically, we found $L_2$ regularization on the difference of neighboring coefficients $\theta_{k+1} - \theta_k$ to perform best).
However, this requires hand-tuning the regularization for every dataset, which defeats the purpose of \emph{learning} the coefficients from the graph. Moreover, we found that even with hand-tuned regularization the coefficients defined by PPR and the heat kernel perform better than trained $\theta_k$, as shown in Fig. \ref{fig:adadif}.

\textbf{How does the label rate affect GDC?} When varying the label rate from 5 up to 60 labels per class we found that the improvement achieved by GDC increases the sparser the labels are. Still, GDC improves performance even for 60 labels per class, i.e. \SI{17}{\percent} label rate (see Fig. \ref{fig:label_rate}). This trend is most likely due to larger neighborhood leveraged by GDC.

\begin{figure}
    \centering
    \input{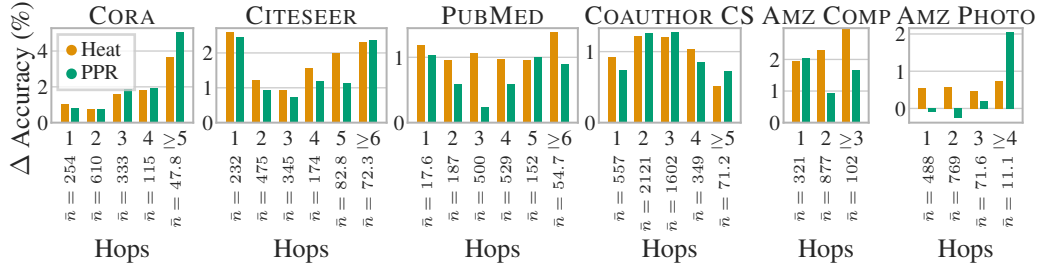}
    \caption{Improvement (percentage points) in GCN accuracy by adding GDC depending on distance (number of hops) from the training set. Nodes further away tend to benefit more from GDC.}
    \label{fig:dist}
\end{figure}

\textbf{Which nodes benefit from GDC?} Our experiments showed no correlation of improvement with most common node properties, except for the distance from the training set. Nodes further away from the training set tend to benefit more from GDC, as demonstrated by Fig. \ref{fig:dist}. Besides smoothing out the neighborhood, GDC also has the effect of increasing the model's range, since it is no longer restricted to only using first-hop neighbors. Hence, nodes further away from the training set influence the training and later benefit from the improved model weights.

\section{Conclusion} \label{sec:conclusion}

We propose graph diffusion convolution (GDC), a method based on sparsified generalized graph diffusion. GDC is a more powerful, yet spatially localized extension of message passing in GNNs, but able to enhance any graph-based model. We show the tight connection between GDC and spectral-based models and analyzed GDC's spectral properties. GDC shares many of the strengths of spectral methods and none of their weaknesses. We conduct extensive and rigorous experiments that show that GDC consistently improves the accuracy of a wide range of models on both supervised and unsupervised tasks across various homophilic datasets and requires very little hyperparameter tuning. There are many extensions and applications of GDC that remain to be explored. We expect many graph-based models and tasks to benefit from GDC, e.g. graph classification and regression. Promising extensions include other diffusion coefficients $\theta_k$ such as those given by the methods presented in \citet{fouss_experimental_2012} and more advanced random walks and operators that are not defined by powers of a transition matrix.


\subsubsection*{Acknowledgments}

This research was supported by the German Federal Ministry of Education and Research (BMBF), grant no. 01IS18036B, and by the Deutsche Forschungsgemeinschaft (DFG) through the Emmy Noether grant GU 1409/2-1 and the TUM International Graduate School of Science and Engineering (IGSSE), GSC 81. The authors of this work take full responsibilities for its content.

\small
\bibliography{gdc}
\bibliographystyle{gdc}
\normalsize


\appendix


\section{Graph diffusion as a polynomial filter} \label{app:difffilter}

We want to find a direct correspondence between graph diffusion with $\theta_k$ and a polynomial filter with parameters $\xi_j$, i.e.
\begin{equation}
    \sum_{j=0}^{J} \xi_j \mL^j \overset{!}{=} \sum_{k=0}^{K} \theta_k \mT^k.
\end{equation}
To do so, we first expand $\mT = \mI_N - \mL$ and use the binomial equation, i.e.
\begin{equation}
    \begin{aligned}
        \sum_{k=0}^{K} \theta_k \mT^k &= \sum_{k=0}^{K} \theta_k (\mI_N - \mL)^k =\\
        &= \sum_{k=0}^{K} \theta_k \sum_{j=0}^k \binom{k}{j} (-1)^j \mI_N^{k-j} \mL^j =\\
        &= \sum_{k=0}^{K} \sum_{j=0}^k \binom{k}{j} \theta_k (-1)^j \mL^j =\\
        &= \sum_{\substack{j, k \in [0, K]\\j \le k}} \binom{k}{j} \theta_k (-1)^j \mL^j =\\
        &= \sum_{j=0}^{K} \underbrace{\sum_{k=j}^{K} \binom{k}{j} \theta_k (-1)^j}_{\xi_j} \mL^j,
    \end{aligned}
\end{equation}
where we recognize the coefficients $\xi_j$ and see that we need to set $J = K$. Note that we reordered the summation indices by recognizing the triangular sum, i.e. the sum over index pairs $(j, k)$ with $j \le k$. The equation for conversion in the opposite direction is obtained in the same way since $\mL = \mI_N - \mT$. To obtain a more convenient form for $K \to \infty$ we shift the summation index using $m = k - j$, i.e.
\begin{equation}
    \xi_j = \sum_{k=j}^{K} \binom{k}{j} (-1)^j \theta_k = \sum_{m=0}^{K - j} \binom{m+j}{j} (-1)^j \theta_{m+j}.
\end{equation}
To find corresponding coefficients for the heat kernel, we let $K \to \infty$, set $\theta_k = e^{-t} \frac{t^k}{k!}$, and use the exponential series to obtain
\begin{equation}
    \begin{aligned}
        \xi^{\text{HK}}_j &= \sum_{m=0}^{\infty} \binom{m+j}{j} (-1)^j e^{-t} \frac{t^{m+j}}{(m+j)!} =\\
        &= \sum_{m=0}^{\infty} \frac{(m+j)!}{m! j!} (-1)^j e^{-t} \frac{t^{m+j}}{(m+j)!} =\\
        &=e^{-t} \frac{(-t)^j}{j!} \sum_{m=0}^{\infty} \frac{t^m}{m!} = \frac{(-t)^j}{j!} e^{-t} e^t = \frac{(-t)^j}{j!}.
    \end{aligned}
\end{equation}
To obtain the coefficients for PPR, we let $K \to \infty$, set $\theta_k = \alpha (1 - \alpha)^k$, and recognize the series expansion $\frac{1}{(1 - x)^{j+1}} = \sum_{m=0}^{\infty} \binom{m+j}{m} x^m$, resulting in
\begin{equation}
    \begin{aligned}
        \xi^{\text{PPR}}_j &= \sum_{m=0}^{\infty} \binom{m+j}{j} (-1)^j \alpha (1 - \alpha)^{m+j} =\\
        &= \alpha (-1)^j (1 - \alpha)^j \sum_{m=0}^{\infty} \binom{m+j}{m} (1 - \alpha)^m =\\
        &= \alpha (\alpha - 1)^j \frac{1}{\alpha^{j+1}} = \left( 1 - \frac{1}{\alpha} \right)^j.
    \end{aligned}
\end{equation}

\section{Experiments} \label{app:exp}

For optimizing the hyperparameters for node classification the data is split into a development and a test set. The development set contains 1500 nodes for all datasets but for \textsc{Coauthor CS}, where it contains 5000 nodes. All remaining nodes are part of the test set and only used once for testing. The development set is split into a training set containing 20 nodes per class and a validation set with the remaining nodes. For every run the accuracy is determined using 100 different random splits of the development set using fixed seeds. Different seeds are used for validation and test splits.
Early stopping patience is set to 100 epochs with a maximum limit of 10000 epochs, which is never reached. The patience is reset after an increase in accuracy on the validation set. For the test runs we select the hyperparameter configurations that showed the highest average accuracy on the validation splits.

We use the same development set for optimizing the hyperparameters for clustering. The test set is only once for generating test results. Clustering results are averaged over 20 randomly initialized runs.

Confidence intervals are calculated by bootstrapping the accuracy results from 100 or 20 runs, respectively, with 1000 samples. All implementations for node classification as well as DGI are based on PyTorch \citep{paszke_automatic_2017} and PyTorch Geometric \citep{fey_fast_2019}. The remaining experiments are based on NumPy \citep{van_der_walt_numpy_2011}, SciPy \citep{jones_scipy:_2001}, graph-tool \citep{peixoto_graph-tool_2014}, and gensim \citep{rehurek_software_2010}. For $k$-means clustering we use the implementation by scikit-learn \citep{pedregosa_scikit-learn:_2011}. All datasets are included in PyTorch Geometric, available at \url{https://github.com/rusty1s/pytorch_geometric}. Experiments using PyTorch are run on Nvidia GPUs using CUDA and the remaining experiments are run on Intel CPUs.

For all experiments the largest connected component of the graph is selected. Dropout probability is set to $p = 0.5$ for all experiments and performed after every application of the activation function. PPR preprocessing is done with $\alpha \in [0.05, 0.30]$, heat kernel preprocessing with $t \in [1, 10]$. For top-$k$ matrix sparsification $k$ is set to either 64 or 128 and for $\epsilon$-thresholding $\epsilon$ is chosen from $[0.00001, 0.01]$. We do not choose $\epsilon$ directly but rather calculate which $\epsilon$ corresponds to a chosen average degree.
For node classification we use the Adam optimizer with a learning rate of 0.01. The hidden dimension of GNNs is kept fixed at 64 with the exception of ARMA, where the dimensionality of a single stack is chosen from 16 or 32. For ARMA, up to three stacks and two layers are tested. GCN and GAT are run with up to 4 layers, JK and GIN with up to six layers. $L_\text{2}$-regularization is performed on the parameters of the first layer of every model with $\lambda_{L_\text{1}} \in [0.001, 10]$.
Unsupervised models use a node embedding dimension of 128. DGI uses the Adam optimizer with a learning rate of 0.001.
For a full list of final hyperparameters per model, diffusion, and dataset see Sec. \ref{sec:hyppar}.

\FloatBarrier
\subsection{Datasets}

\begin{table}[h]
    \centering
    \caption{Dataset statistics.}
    \label{tab:datasets}
    \resizebox{\textwidth}{!}{
        \begin{tabular}{l l r r r r r r}
            Dataset & Type & Classes & Features & Nodes & Edges & Label rate\\
            \hline
            \textsc{Cora} & Citation & \num{7} & \num{1433} & \num{2485} & \num{5069} & \num{0.056}\\
            \textsc{Citeseer} & Citation & \num{6} & \num{3703} & \num{2120} & \num{3679} & \num{0.057}\\
            \textsc{PubMed} & Citation & \num{3} & \num{500} & \num{19717} & \num{44324} & \num{0.003}\\
            \textsc{Coauthor CS} & Co-author & \num{15} & \num{6805} & \num{18333} & \num{81894} & \num{0.016}\\
            \textsc{Amz Comp} & Co-purchase & \num{10} & \num{767} & \num{13381} & \num{245778} & \num{0.015}\\
            \textsc{Amz Photos} & Co-purchase & \num{8} & \num{745} & \num{7487} & \num{119043} & \num{0.021}\\
        \end{tabular}
    }
\end{table}

\FloatBarrier
\subsection{Results} \label{app:results}

To support our claim of achieving state-of-the-art node classification performance we also include results (and hyperparameters) of APPNP, which has been shown to be the current state of the art for semi-supervised node classification and uses graph diffusion internally \citep{gasteiger_predict_2019,fey_fast_2019}.

\begin{table}[h]
    \centering
    \caption{Average accuracy (\%) on \textsc{Cora} with bootstrap-estimated 95\% confidence levels.}
    \label{tab:hyper:cora}
    \resizebox{
        \ifdim\width>\textwidth
            \textwidth
        \else
            \width
        \fi
    }{!}{
\begin{tabular}{lcccc}

    Model &                No diffusion &               Heat &                         PPR &             AdaDIF \\
\hline\\[-0.7em]
      GCN &           \num{81.71+-0.26} &  \num{83.48+-0.22} &           \num{83.58+-0.23} &  \num{82.93+-0.23} \\
      GAT &           \num{80.10+-0.34} &  \num{81.54+-0.25} &           \num{81.60+-0.25} &  \num{81.32+-0.22} \\
       JK &           \num{82.14+-0.24} &  \num{83.69+-0.29} &           \num{83.78+-0.22} &  \num{83.43+-0.21} \\
      GIN &           \num{73.96+-0.46} &  \num{76.54+-0.63} &           \num{78.74+-0.44} &  \num{75.94+-0.45} \\
     ARMA &           \num{81.62+-0.24} &  \num{83.32+-0.22} &           \num{83.81+-0.21} &  \num{83.24+-0.22} \\
    APPNP &  \textbf{\num{83.83+-0.23}} &                  - &                           - &                  - \\
\hline\\[-0.7em]
    DCSBM &           \num{59.75+-1.59} &  \num{64.63+-2.60} &           \num{68.52+-1.47} &                  - \\
 Spectral &           \num{29.29+-1.03} &  \num{35.16+-2.96} &           \num{34.03+-2.01} &                  - \\
 DeepWalk &           \num{68.67+-1.01} &  \num{68.76+-0.67} &           \num{69.42+-0.07} &                  - \\
      DGI &           \num{54.29+-1.21} &  \num{67.71+-1.69} &  \textbf{\num{69.61+-1.73}} &                  - \\

\end{tabular}
    }
\end{table}
\begin{table}[h]
    \centering
    \caption{Average accuracy (\%) on \textsc{Citeseer} with bootstrap-estimated 95\% confidence levels.}
    \label{tab:hyper:citeseer}
    \resizebox{
        \ifdim\width>\textwidth
            \textwidth
        \else
            \width
        \fi
    }{!}{
\begin{tabular}{lcccc}

    Model &       No diffusion &               Heat &                         PPR &             AdaDIF \\
\hline\\[-0.7em]
      GCN &  \num{72.02+-0.31} &  \num{73.22+-0.27} &  \textbf{\num{73.35+-0.27}} &  \num{71.58+-0.31} \\
      GAT &  \num{69.52+-0.32} &  \num{70.25+-0.34} &           \num{68.50+-0.21} &  \num{68.68+-0.22} \\
       JK &  \num{70.34+-0.38} &  \num{72.38+-0.27} &           \num{72.24+-0.31} &  \num{71.11+-0.33} \\
      GIN &  \num{61.09+-0.58} &  \num{62.82+-0.50} &           \num{64.07+-0.48} &  \num{61.46+-0.51} \\
     ARMA &  \num{70.84+-0.32} &  \num{71.90+-0.33} &           \num{72.28+-0.29} &  \num{71.45+-0.31} \\
    APPNP &  \num{72.76+-0.25} &                  - &                           - &                  - \\
\hline\\[-0.7em]
    DCSBM &  \num{46.70+-2.18} &  \num{56.81+-1.21} &           \num{57.14+-1.40} &                  - \\
 Spectral &  \num{27.02+-0.57} &  \num{29.61+-1.29} &           \num{29.26+-1.46} &                  - \\
 DeepWalk &  \num{55.33+-1.05} &  \num{66.05+-0.56} &           \num{65.81+-0.16} &                  - \\
      DGI &  \num{54.62+-2.28} &  \num{71.58+-0.94} &  \textbf{\num{72.42+-0.39}} &                  - \\

\end{tabular}
    }
\end{table}

\begin{table}[h]
    \centering
    \caption{Average accuracy (\%) on \textsc{PubMed} with bootstrap-estimated 95\% confidence levels.}
    \label{tab:hyper:pubmed}
    \resizebox{
        \ifdim\width>\textwidth
            \textwidth
        \else
            \width
        \fi
    }{!}{
\begin{tabular}{lcccc}

    Model &       No diffusion &                        Heat &                PPR &             AdaDIF \\
\hline\\[-0.7em]
      GCN &  \num{78.23+-0.40} &           \num{79.62+-0.36} &  \num{78.72+-0.37} &  \num{77.46+-0.36} \\
      GAT &  \num{76.32+-0.47} &           \num{77.78+-0.34} &  \num{76.66+-0.32} &  \num{75.98+-0.33} \\
       JK &  \num{78.47+-0.36} &  \textbf{\num{79.95+-0.28}} &  \num{79.22+-0.32} &  \num{78.01+-0.41} \\
      GIN &  \num{72.38+-0.63} &           \num{74.16+-0.62} &  \num{73.62+-0.63} &  \num{68.14+-0.80} \\
     ARMA &  \num{77.14+-0.36} &           \num{79.64+-0.35} &  \num{78.85+-0.36} &  \num{77.32+-0.37} \\
    APPNP &  \num{79.78+-0.33} &                           - &                  - &                  - \\
\hline\\[-0.7em]
    DCSBM &  \num{46.64+-1.85} &           \num{67.38+-1.45} &  \num{64.51+-1.75} &                  - \\
 Spectral &  \num{37.97+-0.02} &           \num{49.28+-3.08} &  \num{48.05+-2.69} &                  - \\
 DeepWalk &  \num{70.77+-0.14} &  \textbf{\num{71.36+-0.14}} &  \num{69.96+-0.12} &                  - \\
      DGI &  \num{49.96+-2.21} &           \num{65.94+-0.23} &  \num{66.52+-0.35} &                  - \\

\end{tabular}
    }
\end{table}
\begin{table}[h]
    \centering
    \caption{Average accuracy (\%) on \textsc{Coauthor CS} with bootstrap-estimated 95\% confidence levels.}
    \label{tab:hyper:coauthorcs}
    \resizebox{
        \ifdim\width>\textwidth
            \textwidth
        \else
            \width
        \fi
    }{!}{
\begin{tabular}{lcccc}

    Model &       No diffusion &               Heat &                         PPR &             AdaDIF \\
\hline\\[-0.7em]
      GCN &  \num{91.83+-0.08} &  \num{92.79+-0.07} &  \textbf{\num{93.01+-0.07}} &  \num{92.28+-0.06} \\
      GAT &  \num{90.89+-0.13} &  \num{89.82+-0.10} &           \num{91.33+-0.07} &  \num{88.29+-0.06} \\
       JK &  \num{91.11+-0.09} &  \num{92.40+-0.08} &           \num{92.41+-0.07} &  \num{91.68+-0.08} \\
     ARMA &  \num{91.32+-0.08} &  \num{92.32+-0.09} &           \num{92.63+-0.08} &  \num{91.03+-0.09} \\
    APPNP &  \num{92.08+-0.07} &                  - &                           - &                  - \\
\hline\\[-0.7em]
    DCSBM &  \num{57.70+-1.52} &  \num{63.70+-0.93} &           \num{61.71+-1.15} &                  - \\
 Spectral &  \num{24.74+-2.28} &  \num{50.47+-3.20} &           \num{55.27+-3.00} &                  - \\
 DeepWalk &  \num{61.26+-0.91} &  \num{63.77+-1.28} &  \textbf{\num{65.29+-1.40}} &                  - \\
      DGI &  \num{57.52+-2.63} &  \num{62.84+-1.84} &           \num{63.79+-1.89} &                  - \\

\end{tabular}
    }
\end{table}
\begin{table}[h]
    \centering
    \caption{Average accuracy (\%) on \textsc{Amz Comp} with bootstrap-estimated 95\% confidence levels.}
    \label{tab:hyper:computers}
    \resizebox{
        \ifdim\width>\textwidth
            \textwidth
        \else
            \width
        \fi
    }{!}{
\begin{tabular}{lcccc}

    Model &       No diffusion &                        Heat &                         PPR &             AdaDIF \\
\hline\\[-0.7em]
      GCN &  \num{84.75+-0.23} &  \textbf{\num{86.77+-0.21}} &           \num{86.04+-0.24} &  \num{85.73+-0.23} \\
      GAT &  \num{45.37+-4.20} &           \num{86.68+-0.26} &           \num{85.37+-0.33} &  \num{86.55+-0.26} \\
       JK &  \num{83.33+-0.27} &           \num{86.51+-0.26} &           \num{85.66+-0.30} &  \num{84.40+-0.32} \\
      GIN &  \num{55.44+-0.83} &           \num{81.11+-0.62} &           \num{75.08+-1.20} &  \num{56.52+-1.65} \\
     ARMA &  \num{84.36+-0.26} &           \num{86.09+-0.27} &           \num{84.92+-0.26} &  \num{84.92+-0.29} \\
    APPNP &  \num{81.72+-0.25} &                           - &                           - &                  - \\
\hline\\[-0.7em]
    DCSBM &  \num{44.61+-0.77} &           \num{55.80+-1.29} &  \textbf{\num{57.92+-2.25}} &                  - \\
 Spectral &  \num{40.39+-1.11} &           \num{50.89+-3.05} &           \num{52.62+-2.14} &                  - \\
 DeepWalk &  \num{55.61+-0.25} &           \num{56.29+-0.50} &           \num{55.05+-0.98} &                  - \\
      DGI &  \num{30.84+-1.96} &           \num{37.27+-1.21} &           \num{36.81+-1.12} &                  - \\

\end{tabular}
    }
\end{table}
\begin{table}[h]
    \centering
    \caption{Average accuracy (\%) on \textsc{Amz Photo} with bootstrap-estimated 95\% confidence levels.}
    \label{tab:hyper:photo}
    \resizebox{
        \ifdim\width>\textwidth
            \textwidth
        \else
            \width
        \fi
    }{!}{
\begin{tabular}{lcccc}

    Model &       No diffusion &                        Heat &                PPR &             AdaDIF \\
\hline\\[-0.7em]
      GCN &  \num{92.08+-0.20} &           \num{92.82+-0.23} &  \num{92.20+-0.22} &  \num{92.37+-0.22} \\
      GAT &  \num{53.40+-5.49} &           \num{91.86+-0.20} &  \num{90.89+-0.27} &  \num{91.65+-0.20} \\
       JK &  \num{91.07+-0.26} &  \textbf{\num{92.93+-0.21}} &  \num{92.37+-0.22} &  \num{92.34+-0.22} \\
      GIN &  \num{68.34+-1.16} &           \num{87.24+-0.65} &  \num{83.41+-0.82} &  \num{75.37+-0.86} \\
     ARMA &  \num{91.41+-0.22} &           \num{92.05+-0.24} &  \num{91.09+-0.24} &  \num{90.38+-0.28} \\
    APPNP &  \num{91.42+-0.26} &                           - &                  - &                  - \\
\hline\\[-0.7em]
    DCSBM &  \num{66.30+-1.70} &           \num{67.13+-2.49} &  \num{64.28+-1.81} &                  - \\
 Spectral &  \num{28.15+-0.81} &           \num{49.86+-2.06} &  \num{53.65+-3.22} &                  - \\
 DeepWalk &  \num{78.82+-0.85} &  \textbf{\num{79.26+-0.09}} &  \num{78.73+-0.10} &                  - \\
      DGI &  \num{40.09+-2.14} &           \num{49.02+-1.78} &  \num{51.34+-1.96} &                  - \\

\end{tabular}
    }
\end{table}


\begin{figure}[h]
    \centering
    \input{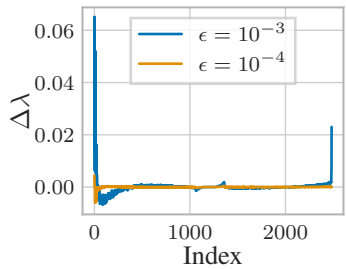}
    \caption{Close-up of difference caused by sparsification (Fig. \ref{fig:filter:sparsif}). Primarily the lowest and highest eigenvalues of the Laplacian are affected.}
    \label{fig:app:sparsif_diff}
\end{figure}

\begin{figure}[h]
    \centering
    \input{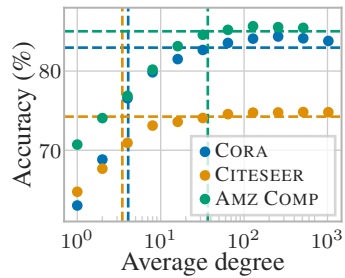}
    \caption{GCN+GDC accuracy (using PPR and sparsification by threshold $\epsilon$). Lines indicate original accuracy and degree. GDC surpasses the original accuracy at around the same degree independent of dataset. Sparsification can improve accuracy.}
    \label{fig:app:degreeeps}
\end{figure}

\FloatBarrier
\subsection{Hyperparameters} \label{sec:hyppar}

\begin{table}[h]
    \centering
    \caption{Hyperparameters for GCN obtained by grid and random search.}
    \label{tab:hyper:gcn}
    \resizebox{
        \ifdim\width>\textwidth
            \textwidth
        \else
            \width
        \fi
    }{!}{
\begin{tabular}{llccccccccc}
Diffusion & Dataset name & $\alpha$ & $t$ & $k$ & $\epsilon$ & $\lambda_{L_\text{2}}$ & \shortstack{Learning\\rate} & Dropout & \shortstack{Hidden\\dimension} & \shortstack{Hidden\\depth} \\
\hline\\[-0.7em]
\multirow6{*}{-} 	& \textsc{Cora} 	& \multirow6{*}{-} 	& \multirow6{*}{-} 	& \multirow6{*}{-} 	& \multirow6{*}{-} 	& 0.06 	& \multirow6{*}{0.01} 	& \multirow6{*}{0.5} 	& \multirow6{*}{64} 	& \multirow6{*}{1} 	\\
		 & \textsc{Citeseer} 	& 		 & 		 & 		 & 		 & 10.0 	& 		 & 		 & 		 & 		 \\
		 & \textsc{PubMed} 	& 		 & 		 & 		 & 		 & 0.03 	& 		 & 		 & 		 & 		 \\
		 & \textsc{Coauthor CS} 	& 		 & 		 & 		 & 		 & 0.06 	& 		 & 		 & 		 & 		 \\
		 & \textsc{Amz Comp} 	& 		 & 		 & 		 & 		 & 0.03 	& 		 & 		 & 		 & 		 \\
		 & \textsc{Amz Photo} 	& 		 & 		 & 		 & 		 & 0.03 	& 		 & 		 & 		 & 		 \\
\hline\\[-0.7em]
\multirow6{*}{Heat} 	& \textsc{Cora} 	& \multirow6{*}{-} 	& 5 	& - 	& 0.0001 	& 0.09 	& \multirow6{*}{0.01} 	& \multirow6{*}{0.5} 	& \multirow6{*}{64} 	& 1 	\\
		 & \textsc{Citeseer} 	& 		 & 4 	& - 	& 0.0009 	& 10.0 	& 		 & 		 & 		 & 1 	\\
		 & \textsc{PubMed} 	& 		 & 3 	& - 	& 0.0001 	& 0.04 	& 		 & 		 & 		 & 1 	\\
		 & \textsc{Coauthor CS} 	& 		 & 1 	& 64 	& - 	& 0.08 	& 		 & 		 & 		 & 1 	\\
		 & \textsc{Amz Comp} 	& 		 & 5 	& - 	& 0.0010 	& 0.07 	& 		 & 		 & 		 & 1 	\\
		 & \textsc{Amz Photo} 	& 		 & 3 	& - 	& 0.0001 	& 0.08 	& 		 & 		 & 		 & 2 	\\
\hline\\[-0.7em]
\multirow6{*}{PPR} 	& \textsc{Cora} 	& 0.05 	& \multirow6{*}{-} 	& 128 	& \multirow6{*}{-} 	& 0.10 	& \multirow6{*}{0.01} 	& \multirow6{*}{0.5} 	& \multirow6{*}{64} 	& \multirow6{*}{1} 	\\
		 & \textsc{Citeseer} 	& 0.10 	& 		 & 	& 0.0009	 & 10.0 	& 		 & 		 & 		 & 		 \\
		 & \textsc{PubMed} 	& 0.10 	& 		 & 64 	& 		 & 0.06 	& 		 & 		 & 		 & 		 \\
		 & \textsc{Coauthor CS} 	& 0.10 	& 		 & 64 	& 		 & 0.03 	& 		 & 		 & 		 & 		 \\
		 & \textsc{Amz Comp} 	& 0.10 	& 		 & 64 	& 		 & 0.04 	& 		 & 		 & 		 & 		 \\
		 & \textsc{Amz Photo} 	& 0.15 	& 		 & 64 	& 		 & 0.03 	& 		 & 		 & 		 & 		 \\
\hline\\[-0.7em]
\multirow6{*}{AdaDIF} 	& \textsc{Cora} 	& \multirow6{*}{-} 	& \multirow6{*}{-} 	& 128 	& \multirow6{*}{-} 	& 0.08 	& \multirow6{*}{0.01} 	& \multirow6{*}{0.5} 	& \multirow6{*}{64} 	& 1 	\\
		 & \textsc{Citeseer} 	& 		 & 		 & 128 	& 		 & 0.08 	& 		 & 		 & 		 & 1 	\\
		 & \textsc{PubMed} 	& 		 & 		 & 128 	& 		 & 0.01 	& 		 & 		 & 		 & 2 	\\
		 & \textsc{Coauthor CS} 	& 		 & 		 & 64 	& 		 & 0.03 	& 		 & 		 & 		 & 1 	\\
		 & \textsc{Amz Comp} 	& 		 & 		 & 64 	& 		 & 0.02 	& 		 & 		 & 		 & 1 	\\
		 & \textsc{Amz Photo} 	& 		 & 		 & 64 	& 		 & 0.02 	& 		 & 		 & 		 & 1 	\\

\end{tabular}

}
\end{table}

\begin{table}[h]
    \centering
    \caption{Hyperparameters for GAT obtained by grid and random search.}
    \label{tab:hyper:gat}
    \resizebox{
        \ifdim\width>\textwidth
            \textwidth
        \else
            \width
        \fi
    }{!}{
\begin{tabular}{llccccccccc}
Diffusion & Dataset name & $\alpha$ & $t$ & $k$ & $\epsilon$ & $\lambda_{L_\text{2}}$ & \shortstack{Learning\\rate} & Dropout & \shortstack{Hidden\\dimension} & \shortstack{Hidden\\depth} \\
\hline\\[-0.7em]
\multirow6{*}{-} 	& \textsc{Cora} 	& \multirow6{*}{-} 	& \multirow6{*}{-} 	& \multirow6{*}{-} 	& \multirow6{*}{-} 	& 0.06 	& \multirow6{*}{0.01} 	& \multirow6{*}{0.5} 	& \multirow6{*}{64} 	& 1 	\\
		 & \textsc{Citeseer} 	& 		 & 		 & 		 & 		 & 0.06 	& 		 & 		 & 		 & 1 	\\
		 & \textsc{PubMed} 	& 		 & 		 & 		 & 		 & 0.03 	& 		 & 		 & 		 & 2 	\\
		 & \textsc{Coauthor CS} 	& 		 & 		 & 		 & 		 & 0.00 	& 		 & 		 & 		 & 2 	\\
		 & \textsc{Amz Comp} 	& 		 & 		 & 		 & 		 & 0.09 	& 		 & 		 & 		 & 1 	\\
		 & \textsc{Amz Photo} 	& 		 & 		 & 		 & 		 & 0.08 	& 		 & 		 & 		 & 1 	\\
\hline\\[-0.7em]
\multirow6{*}{Heat} 	& \textsc{Cora} 	& \multirow6{*}{-} 	& \multirow6{*}{1} 	& \multirow6{*}{-} 	& 0.0010 	& 0.04 	& \multirow6{*}{0.01} 	& \multirow6{*}{0.5} 	& \multirow6{*}{64} 	& 1 	\\
		 & \textsc{Citeseer} 	& 		 & 		 & 		 & 0.0010 	& 0.08 	& 		 & 		 & 		 & 1 	\\
		 & \textsc{PubMed} 	& 		 & 		 & 		 & 0.0005 	& 0.02 	& 		 & 		 & 		 & 2 	\\
		 & \textsc{Coauthor CS} 	& 		 & 		 & 		 & 0.0005 	& 0.03 	& 		 & 		 & 		 & 1 	\\
		 & \textsc{Amz Comp} 	& 		 & 		 & 		 & 0.0005 	& 0.01 	& 		 & 		 & 		 & 1 	\\
		 & \textsc{Amz Photo} 	& 		 & 		 & 		 & 0.0005 	& 0.01 	& 		 & 		 & 		 & 1 	\\
\hline\\[-0.7em]
\multirow6{*}{PPR} 	& \textsc{Cora} 	& \multirow6{*}{0.10} 	& \multirow6{*}{-} 	& \multirow6{*}{-} 	& 0.0050 	& 0.08 	& \multirow6{*}{0.01} 	& \multirow6{*}{0.5} 	& \multirow6{*}{64} 	& 1 	\\
		 & \textsc{Citeseer} 	& 		 & 		 & 		 & 0.0005 	& 0.10 	& 		 & 		 & 		 & 1 	\\
		 & \textsc{PubMed} 	& 		 & 		 & 		 & 0.0005 	& 0.00 	& 		 & 		 & 		 & 2 	\\
		 & \textsc{Coauthor CS} 	& 		 & 		 & 		 & 0.0005 	& 0.00 	& 		 & 		 & 		 & 1 	\\
		 & \textsc{Amz Comp} 	& 		 & 		 & 		 & 0.0005 	& 0.03 	& 		 & 		 & 		 & 1 	\\
		 & \textsc{Amz Photo} 	& 		 & 		 & 		 & 0.0005 	& 0.07 	& 		 & 		 & 		 & 2 	\\
\hline\\[-0.7em]
\multirow6{*}{AdaDIF} 	& \textsc{Cora} 	& \multirow6{*}{-} 	& \multirow6{*}{-} 	& - 	& 0.0010 	& 0.04 	& \multirow6{*}{0.01} 	& \multirow6{*}{0.5} 	& \multirow6{*}{64} 	& 1 	\\
		 & \textsc{Citeseer} 	& 		 & 		 & - 	& 0.0005 	& 0.04 	& 		 & 		 & 		 & 1 	\\
		 & \textsc{PubMed} 	& 		 & 		 & 128 	& - 	& 0.01 	& 		 & 		 & 		 & 2 	\\
		 & \textsc{Coauthor CS} 	& 		 & 		 & 64 	& - 	& 0.02 	& 		 & 		 & 		 & 1 	\\
		 & \textsc{Amz Comp} 	& 		 & 		 & 64 	& - 	& 0.02 	& 		 & 		 & 		 & 1 	\\
		 & \textsc{Amz Photo} 	& 		 & 		 & 64 	& - 	& 0.02 	& 		 & 		 & 		 & 1 	\\

\end{tabular}

}
\end{table}
\begin{table}[h]
    \centering
    \caption{Hyperparameters for JK obtained by grid and random search.}
    \label{tab:hyper:jk}
    \resizebox{
        \ifdim\width>\textwidth
            \textwidth
        \else
            \width
        \fi
    }{!}{
\begin{tabular}{llcccccccccc}
Diffusion & Dataset name & $\alpha$ & $t$ & $k$ & $\epsilon$ & $\lambda_{L_\text{2}}$ & \shortstack{Learning\\rate} & Dropout & Aggregation & \shortstack{Hidden\\dimension} & \shortstack{Hidden\\depth} \\
\hline\\[-0.7em]
\multirow6{*}{-} 	& \textsc{Cora} 	& \multirow6{*}{-} 	& \multirow6{*}{-} 	& \multirow6{*}{-} 	& \multirow6{*}{-} 	& 0.04 	& \multirow6{*}{0.01} 	& \multirow6{*}{0.5} 	& \multirow6{*}{Concatenation} 	& \multirow6{*}{64} 	& 3 	\\
		 & \textsc{Citeseer} 	& 		 & 		 & 		 & 		 & 1.00 	& 		 & 		 & 		 & 		 & 4 	\\
		 & \textsc{PubMed} 	& 		 & 		 & 		 & 		 & 0.05 	& 		 & 		 & 		 & 		 & 2 	\\
		 & \textsc{Coauthor CS} 	& 		 & 		 & 		 & 		 & 0.02 	& 		 & 		 & 		 & 		 & 2 	\\
		 & \textsc{Amz Comp} 	& 		 & 		 & 		 & 		 & 0.03 	& 		 & 		 & 		 & 		 & 2 	\\
		 & \textsc{Amz Photo} 	& 		 & 		 & 		 & 		 & 0.03 	& 		 & 		 & 		 & 		 & 2 	\\
\hline\\[-0.7em]
\multirow6{*}{Heat} 	& \textsc{Cora} 	& \multirow6{*}{-} 	& 5 	& - 	& 0.0001 	& 0.09 	& \multirow6{*}{0.01} 	& \multirow6{*}{0.5} 	& \multirow6{*}{Concatenation} 	& \multirow6{*}{64} 	& \multirow6{*}{2} 	\\
		 & \textsc{Citeseer} 	& 		 & 4 	& - 	& 0.0009 	& 1.00 	& 		 & 		 & 		 & 		 & 		 \\
		 & \textsc{PubMed} 	& 		 & 3 	& - 	& 0.0001 	& 0.09 	& 		 & 		 & 		 & 		 & 		 \\
		 & \textsc{Coauthor CS} 	& 		 & 1 	& 64 	& - 	& 0.03 	& 		 & 		 & 		 & 		 & 		 \\
		 & \textsc{Amz Comp} 	& 		 & 5 	& - 	& 0.0010 	& 0.07 	& 		 & 		 & 		 & 		 & 		 \\
		 & \textsc{Amz Photo} 	& 		 & 3 	& - 	& 0.0005 	& 0.07 	& 		 & 		 & 		 & 		 & 		 \\
\hline\\[-0.7em]
\multirow6{*}{PPR} 	& \textsc{Cora} 	& 0.05 	& \multirow6{*}{-} 	& 128 	& \multirow6{*}{-} 	& 0.10 	& \multirow6{*}{0.01} 	& \multirow6{*}{0.5} 	& \multirow6{*}{Concatenation} 	& \multirow6{*}{64} 	& \multirow6{*}{2} 	\\
		 & \textsc{Citeseer} 	& 0.2 	& 		 & 	& 	0.0009 & 1.00 	& 		 & 		 & 		 & 		 & 		 \\
		 & \textsc{PubMed} 	& 0.10 	& 		 & 64 	& 		 & 0.02 	& 		 & 		 & 		 & 		 & 		 \\
		 & \textsc{Coauthor CS} 	& 0.10 	& 		 & 64 	& 		 & 0.03 	& 		 & 		 & 		 & 		 & 		 \\
		 & \textsc{Amz Comp} 	& 0.10 	& 		 & 64 	& 		 & 0.04 	& 		 & 		 & 		 & 		 & 		 \\
		 & \textsc{Amz Photo} 	& 0.15 	& 		 & 64 	& 		 & 0.03 	& 		 & 		 & 		 & 		 & 		 \\
\hline\\[-0.7em]
\multirow6{*}{AdaDIF} 	& \textsc{Cora} 	& \multirow6{*}{-} 	& \multirow6{*}{-} 	& 128 	& \multirow6{*}{-} 	& 0.05 	& \multirow6{*}{0.01} 	& \multirow6{*}{0.5} 	& \multirow6{*}{Concatenation} 	& \multirow6{*}{64} 	& 2 	\\
		 & \textsc{Citeseer} 	& 		 & 		 & 128 	& 		 & 0.08 	& 		 & 		 & 		 & 		 & 2 	\\
		 & \textsc{PubMed} 	& 		 & 		 & 128 	& 		 & 0.01 	& 		 & 		 & 		 & 		 & 3 	\\
		 & \textsc{Coauthor CS} 	& 		 & 		 & 64 	& 		 & 0.02 	& 		 & 		 & 		 & 		 & 2 	\\
		 & \textsc{Amz Comp} 	& 		 & 		 & 64 	& 		 & 0.03 	& 		 & 		 & 		 & 		 & 2 	\\
		 & \textsc{Amz Photo} 	& 		 & 		 & 64 	& 		 & 0.02 	& 		 & 		 & 		 & 		 & 2 	\\

\end{tabular}

}
\end{table}

\begin{table}[h]
    \centering
    \caption{Hyperparameters for GIN obtained by grid and random search.}
    \label{tab:hyper:gin}
    \resizebox{
        \ifdim\width>\textwidth
            \textwidth
        \else
            \width
        \fi
    }{!}{
\begin{tabular}{llcccccccccc}
Diffusion & Dataset name & $\alpha$ & $t$ & $k$ & $\epsilon$ & $\lambda_{L_\text{2}}$ & \shortstack{Learning\\rate} & Dropout & Aggregation & \shortstack{Hidden\\dimension} & \shortstack{Hidden\\depth} \\
\hline\\[-0.7em]
\multirow5{*}{-} 	& \textsc{Cora} 	& \multirow5{*}{-} 	& \multirow5{*}{-} 	& \multirow5{*}{-} 	& \multirow5{*}{-} 	& 0.09 	& \multirow5{*}{0.01} 	& \multirow5{*}{0.5} 	& \multirow5{*}{Sum} 	& \multirow5{*}{64} 	& 4 	\\
		 & \textsc{Citeseer} 	& 		 & 		 & 		 & 		 & 0.10 	& 		 & 		 & 		 & 		 & 4 	\\
		 & \textsc{PubMed} 	& 		 & 		 & 		 & 		 & 0.08 	& 		 & 		 & 		 & 		 & 4 	\\
		 & \textsc{Amz Comp} 	& 		 & 		 & 		 & 		 & 0.01 	& 		 & 		 & 		 & 		 & 5 	\\
		 & \textsc{Amz Photo} 	& 		 & 		 & 		 & 		 & 0.01 	& 		 & 		 & 		 & 		 & 4 	\\
\hline\\[-0.7em]
\multirow5{*}{Heat} 	& \textsc{Cora} 	& \multirow5{*}{-} 	& 3 	& - 	& 0.0001 	& 0.07 	& \multirow5{*}{0.01} 	& \multirow5{*}{0.5} 	& \multirow5{*}{Sum} 	& \multirow5{*}{64} 	& 5 	\\
		 & \textsc{Citeseer} 	& 		 & 8 	& - 	& 0.0009 	& 0.01 	& 		 & 		 & 		 & 		 & 4 	\\
		 & \textsc{PubMed} 	& 		 & 3 	& - 	& 0.0010 	& 0.02 	& 		 & 		 & 		 & 		 & 5 	\\
		 & \textsc{Amz Comp} 	& 		 & 3 	& 64 	& - 	& 0.00 	& 		 & 		 & 		 & 		 & 4 	\\
		 & \textsc{Amz Photo} 	& 		 & 3 	& 64 	& - 	& 0.00 	& 		 & 		 & 		 & 		 & 4 	\\
\hline\\[-0.7em]
\multirow5{*}{PPR} 	& \textsc{Cora} 	& 0.05 	& \multirow5{*}{-} 	& 128 	& \multirow5{*}{-} 	& 0.01 	& \multirow5{*}{0.01} 	& \multirow5{*}{0.5} 	& \multirow5{*}{Sum} 	& \multirow5{*}{64} 	& 4 	\\
		 & \textsc{Citeseer} 	& 0.05 	& 		 & 	& 0.0009	 & 0.01 	& 		 & 		 & 		 & 		 & 4 	\\
		 & \textsc{PubMed} 	& 0.10 	& 		 & 64 	& 		 & 0.01 	& 		 & 		 & 		 & 		 & 5 	\\
		 & \textsc{Amz Comp} 	& 0.10 	& 		 & 64 	& 		 & 0.04 	& 		 & 		 & 		 & 		 & 4 	\\
		 & \textsc{Amz Photo} 	& 0.10 	& 		 & 64 	& 		 & 0.04 	& 		 & 		 & 		 & 		 & 4 	\\
\hline\\[-0.7em]
\multirow5{*}{AdaDIF} 	& \textsc{Cora} 	& \multirow5{*}{-} 	& \multirow5{*}{-} 	& 128 	& \multirow5{*}{-} 	& 0.02 	& \multirow5{*}{0.01} 	& \multirow5{*}{0.5} 	& \multirow5{*}{Sum} 	& \multirow5{*}{64} 	& 3 	\\
		 & \textsc{Citeseer} 	& 		 & 		 & 128 	& 		 & 0.05 	& 		 & 		 & 		 & 		 & 4 	\\
		 & \textsc{PubMed} 	& 		 & 		 & 64 	& 		 & 0.03 	& 		 & 		 & 		 & 		 & 5 	\\
		 & \textsc{Amz Comp} 	& 		 & 		 & 64 	& 		 & 0.02 	& 		 & 		 & 		 & 		 & 4 	\\
		 & \textsc{Amz Photo} 	& 		 & 		 & 64 	& 		 & 0.02 	& 		 & 		 & 		 & 		 & 4 	\\

\end{tabular}

}
\end{table}

\begin{table}[h]
    \centering
    \caption{Hyperparameters for ARMA obtained by grid and random search.}
    \label{tab:hyper:arma}
    \resizebox{
        \ifdim\width>\textwidth
            \textwidth
        \else
            \width
        \fi
    }{!}{
\begin{tabular}{llccccccccccc}
Diffusion & Dataset name & $\alpha$ & $t$ & $k$ & $\epsilon$ & $\lambda_{L_\text{2}}$ & \shortstack{Learning\\rate} & Dropout & \shortstack{ARMA\\layers} & \shortstack{ARMA\\stacks} & \shortstack{Hidden\\dimension} & \shortstack{Hidden\\depth} \\
\hline\\[-0.7em]
\multirow6{*}{-} 	& \textsc{Cora} 	& \multirow6{*}{-} 	& \multirow6{*}{-} 	& \multirow6{*}{-} 	& \multirow6{*}{-} 	& 0.04 	& \multirow6{*}{0.01} 	& \multirow6{*}{0.5} 	& \multirow6{*}{1} 	& 3 	& \multirow6{*}{16} 	& \multirow6{*}{1} 	\\
		 & \textsc{Citeseer} 	& 		 & 		 & 		 & 		 & 0.08 	& 		 & 		 & 		 & 3 	& 		 & 		 \\
		 & \textsc{PubMed} 	& 		 & 		 & 		 & 		 & 0.00 	& 		 & 		 & 		 & 2 	& 		 & 		 \\
		 & \textsc{Coauthor CS} 	& 		 & 		 & 		 & 		 & 0.02 	& 		 & 		 & 		 & 2 	& 		 & 		 \\
		 & \textsc{Amz Comp} 	& 		 & 		 & 		 & 		 & 0.01 	& 		 & 		 & 		 & 3 	& 		 & 		 \\
		 & \textsc{Amz Photo} 	& 		 & 		 & 		 & 		 & 0.01 	& 		 & 		 & 		 & 3 	& 		 & 		 \\
\hline\\[-0.7em]
\multirow6{*}{Heat} 	& \textsc{Cora} 	& \multirow6{*}{-} 	& 5 	& 64 	& - 	& 0.08 	& \multirow6{*}{0.01} 	& \multirow6{*}{0.5} 	& \multirow6{*}{1} 	& 2 	& \multirow6{*}{16} 	& \multirow6{*}{1} 	\\
		 & \textsc{Citeseer} 	& 		 & 5 	& 64 	& - 	& 0.08 	& 		 & 		 & 		 & 3 	& 		 & 		 \\
		 & \textsc{PubMed} 	& 		 & 3 	& - 	& 0.0001 	& 0.00 	& 		 & 		 & 		 & 2 	& 		 & 		 \\
		 & \textsc{Coauthor CS} 	& 		 & 1 	& 64 	& - 	& 0.01 	& 		 & 		 & 		 & 3 	& 		 & 		 \\
		 & \textsc{Amz Comp} 	& 		 & 5 	& 64 	& - 	& 0.04 	& 		 & 		 & 		 & 3 	& 		 & 		 \\
		 & \textsc{Amz Photo} 	& 		 & 3 	& 64 	& - 	& 0.04 	& 		 & 		 & 		 & 2 	& 		 & 		 \\
\hline\\[-0.7em]
\multirow6{*}{PPR} 	& \textsc{Cora} 	& 0.10 	& \multirow6{*}{-} 	& 128 	& \multirow6{*}{-} 	& 0.05 	& \multirow6{*}{0.01} 	& \multirow6{*}{0.5} 	& \multirow6{*}{1} 	& 3 	& 16 	& \multirow6{*}{1} 	\\
		 & \textsc{Citeseer} 	& 0.15 	& 		 & 128 	& 		 & 0.08 	& 		 & 		 & 		 & 3 	& 16 	& 		 \\
		 & \textsc{PubMed} 	& 0.10 	& 		 & 64 	& 		 & 0.01 	& 		 & 		 & 		 & 3 	& 16 	& 		 \\
		 & \textsc{Coauthor CS} 	& 0.10 	& 		 & 64 	& 		 & 0.01 	& 		 & 		 & 		 & 2 	& 16 	& 		 \\
		 & \textsc{Amz Comp} 	& 0.10 	& 		 & 128 	& 		 & 0.06 	& 		 & 		 & 		 & 2 	& 32 	& 		 \\
		 & \textsc{Amz Photo} 	& 0.15 	& 		 & 128 	& 		 & 0.06 	& 		 & 		 & 		 & 2 	& 32 	& 		 \\
\hline\\[-0.7em]
\multirow6{*}{AdaDIF} 	& \textsc{Cora} 	& \multirow6{*}{-} 	& \multirow6{*}{-} 	& 128 	& \multirow6{*}{-} 	& 0.05 	& \multirow6{*}{0.01} 	& \multirow6{*}{0.5} 	& \multirow6{*}{1} 	& 2 	& \multirow6{*}{16} 	& \multirow6{*}{1} 	\\
		 & \textsc{Citeseer} 	& 		 & 		 & 128 	& 		 & 0.09 	& 		 & 		 & 		 & 3 	& 		 & 		 \\
		 & \textsc{PubMed} 	& 		 & 		 & 64 	& 		 & 0.01 	& 		 & 		 & 		 & 2 	& 		 & 		 \\
		 & \textsc{Coauthor CS} 	& 		 & 		 & 64 	& 		 & 0.03 	& 		 & 		 & 		 & 2 	& 		 & 		 \\
		 & \textsc{Amz Comp} 	& 		 & 		 & 64 	& 		 & 0.01 	& 		 & 		 & 		 & 3 	& 		 & 		 \\
		 & \textsc{Amz Photo} 	& 		 & 		 & 64 	& 		 & 0.01 	& 		 & 		 & 		 & 2 	& 		 & 		 \\

\end{tabular}

}
\end{table}
\begin{table}[h]
    \centering
    \caption{Hyperparameters for APPNP obtained by grid and random search.}
    \label{tab:hyper:appnp}
    \resizebox{
        \ifdim\width>\textwidth
            \textwidth
        \else
            \width
        \fi
    }{!}{
\begin{tabular}{llcccccc}
Dataset name & $\alpha$ & $k$ & $\lambda_{L_\text{2}}$ & \shortstack{Learning\\rate} & Dropout & \shortstack{Hidden\\dimension} & \shortstack{Hidden\\depth} \\
\hline\\[-0.7em]
\textsc{Cora} 	& 0.10 	& \multirow6{*}{10} 	& 0.09 	& \multirow6{*}{0.01} 	& \multirow6{*}{0.5} 	& \multirow6{*}{64} 	& \multirow6{*}{1} 	\\
\textsc{Citeseer} 	& 0.10 	& 		 & 1.00 	& 		 & 		 & 		 & 		 \\
\textsc{PubMed} 	& 0.10 	& 		 & 0.02 	& 		 & 		 & 		 & 		 \\
\textsc{Coauthor CS} 	& 0.15 	& 		 & 0.01 	& 		 & 		 & 		 & 		 \\
\textsc{Amz Comp} 	& 0.10 	& 		 & 0.06 	& 		 & 		 & 		 & 		 \\
\textsc{Amz Photo} 	& 0.10 	& 		 & 0.05 	& 		 & 		 & 		 & 		 \\
\hline\\[-0.7em]

\end{tabular}

}
\end{table}

\begin{table}[h]
    \centering
    \caption{Hyperparameters for DCSBM obtained by grid and random search.}
    \label{tab:hyper:dcsbm}
    \resizebox{
        \ifdim\width>\textwidth
            \textwidth
        \else
            \width
        \fi
    }{!}{
\begin{tabular}{llccccc}
Diffusion & Dataset name & $\alpha$ & $t$ & $k$ & $\epsilon$ & Number of blocks \\
\hline\\[-0.7em]
\multirow6{*}{-} 	& \textsc{Cora} 	& \multirow6{*}{-} 	& \multirow6{*}{-} 	& \multirow6{*}{-} 	& \multirow6{*}{-} 	& 7 	\\
		 & \textsc{Citeseer} 	& 		 & 		 & 		 & 		 & 6 	\\
		 & \textsc{PubMed} 	& 		 & 		 & 		 & 		 & 3 	\\
		 & \textsc{Coauthor CS} 	& 		 & 		 & 		 & 		 & 15 	\\
		 & \textsc{Amz Comp} 	& 		 & 		 & 		 & 		 & 10 	\\
		 & \textsc{Amz Photo} 	& 		 & 		 & 		 & 		 & 8 	\\
\hline\\[-0.7em]
\multirow6{*}{Heat} 	& \textsc{Cora} 	& \multirow6{*}{-} 	& 5 	& - 	& 0.0010 	& 7 	\\
		 & \textsc{Citeseer} 	& 		 & 1 	& 64 	& - 	& 6 	\\
		 & \textsc{PubMed} 	& 		 & 3 	& 64 	& - 	& 3 	\\
		 & \textsc{Coauthor CS} 	& 		 & 5 	& - 	& 0.0010 	& 15 	\\
		 & \textsc{Amz Comp} 	& 		 & 3 	& - 	& 0.0010 	& 10 	\\
		 & \textsc{Amz Photo} 	& 		 & 3 	& - 	& 0.0010 	& 8 	\\
\hline\\[-0.7em]
\multirow6{*}{PPR} 	& \textsc{Cora} 	& 0.05 	& \multirow6{*}{-} 	& - 	& 0.0010 	& 7 	\\
		 & \textsc{Citeseer} 	& 0.05 	& 		 & 64 	& - 	& 6 	\\
		 & \textsc{PubMed} 	& 0.10 	& 		 & - 	& 0.0010 	& 3 	\\
		 & \textsc{Coauthor CS} 	& 0.05 	& 		 & 64 	& - 	& 15 	\\
		 & \textsc{Amz Comp} 	& 0.05 	& 		 & - 	& 0.0010 	& 10 	\\
		 & \textsc{Amz Photo} 	& 0.10 	& 		 & 64 	& - 	& 8 	\\
\hline\\[-0.7em]

\end{tabular}

}
\end{table}
\begin{table}[h]
    \centering
    \caption{Hyperparameters for spectral clustering obtained by grid and random search.}
    \label{tab:hyper:spectral}
    \resizebox{
        \ifdim\width>\textwidth
            \textwidth
        \else
            \width
        \fi
    }{!}{
\begin{tabular}{llccccc}
Diffusion & Dataset name & $\alpha$ & $t$ & $k$ & $\epsilon$ & \shortstack{Embedding\\dimension} \\
\hline\\[-0.7em]
\multirow6{*}{-} 	& \textsc{Cora} 	& \multirow6{*}{-} 	& \multirow6{*}{-} 	& \multirow6{*}{-} 	& \multirow6{*}{-} 	& \multirow6{*}{128} 	\\
		 & \textsc{Citeseer} 	& 		 & 		 & 		 & 		 & 		 \\
		 & \textsc{PubMed} 	& 		 & 		 & 		 & 		 & 		 \\
		 & \textsc{Coauthor CS} 	& 		 & 		 & 		 & 		 & 		 \\
		 & \textsc{Amz Comp} 	& 		 & 		 & 		 & 		 & 		 \\
		 & \textsc{Amz Photo} 	& 		 & 		 & 		 & 		 & 		 \\
\hline\\[-0.7em]
\multirow6{*}{Heat} 	& \textsc{Cora} 	& \multirow6{*}{-} 	& 5 	& - 	& 0.0010 	& \multirow6{*}{128} 	\\
		 & \textsc{Citeseer} 	& 		 & 5 	& - 	& 0.0010 	& 		 \\
		 & \textsc{PubMed} 	& 		 & 5 	& 64 	& - 	& 		 \\
		 & \textsc{Coauthor CS} 	& 		 & 5 	& - 	& 0.0010 	& 		 \\
		 & \textsc{Amz Comp} 	& 		 & 1 	& 64 	& - 	& 		 \\
		 & \textsc{Amz Photo} 	& 		 & 5 	& 64 	& - 	& 		 \\
\hline\\[-0.7em]
\multirow6{*}{PPR} 	& \textsc{Cora} 	& 0.10 	& \multirow6{*}{-} 	& - 	& 0.0010 	& \multirow6{*}{128} 	\\
		 & \textsc{Citeseer} 	& 0.05 	& 		 & - 	& 0.0010 	& 		 \\
		 & \textsc{PubMed} 	& 0.15 	& 		 & - 	& 0.0010 	& 		 \\
		 & \textsc{Coauthor CS} 	& 0.05 	& 		 & 64 	& - 	& 		 \\
		 & \textsc{Amz Comp} 	& 0.05 	& 		 & 64 	& - 	& 		 \\
		 & \textsc{Amz Photo} 	& 0.15 	& 		 & 64 	& - 	& 		 \\
\hline\\[-0.7em]

\end{tabular}

}
\end{table}
\begin{table}[h]
    \centering
    \caption{Hyperparameters for DeepWalk obtained by grid and random search.}
    \label{tab:hyper:deepwalk}
    \resizebox{
        \ifdim\width>\textwidth
            \textwidth
        \else
            \width
        \fi
    }{!}{
\begin{tabular}{llccccccc}
Diffusion & Dataset name & $\alpha$ & $t$ & $k$ & $\epsilon$ & Walks per node & \shortstack{Embedding\\dimension} & \shortstack{Walk\\length} \\
\hline\\[-0.7em]
\multirow6{*}{-} 	& \textsc{Cora} 	& \multirow6{*}{-} 	& \multirow6{*}{-} 	& \multirow6{*}{-} 	& \multirow6{*}{-} 	& \multirow6{*}{10} 	& \multirow6{*}{128} 	& \multirow6{*}{64} 	\\
		 & \textsc{Citeseer} 	& 		 & 		 & 		 & 		 & 		 & 		 & 		 \\
		 & \textsc{PubMed} 	& 		 & 		 & 		 & 		 & 		 & 		 & 		 \\
		 & \textsc{Coauthor CS} 	& 		 & 		 & 		 & 		 & 		 & 		 & 		 \\
		 & \textsc{Amz Comp} 	& 		 & 		 & 		 & 		 & 		 & 		 & 		 \\
		 & \textsc{Amz Photo} 	& 		 & 		 & 		 & 		 & 		 & 		 & 		 \\
\hline\\[-0.7em]
\multirow6{*}{Heat} 	& \textsc{Cora} 	& \multirow6{*}{-} 	& 5 	& - 	& 0.0010 	& \multirow6{*}{10} 	& \multirow6{*}{128} 	& \multirow6{*}{64} 	\\
		 & \textsc{Citeseer} 	& 		 & 1 	& 64 	& - 	& 		 & 		 & 		 \\
		 & \textsc{PubMed} 	& 		 & 1 	& - 	& 0.0010 	& 		 & 		 & 		 \\
		 & \textsc{Coauthor CS} 	& 		 & 5 	& - 	& 0.0010 	& 		 & 		 & 		 \\
		 & \textsc{Amz Comp} 	& 		 & 3 	& - 	& 0.0010 	& 		 & 		 & 		 \\
		 & \textsc{Amz Photo} 	& 		 & 3 	& - 	& 0.0010 	& 		 & 		 & 		 \\
\hline\\[-0.7em]
\multirow6{*}{PPR} 	& \textsc{Cora} 	& 0.05 	& \multirow6{*}{-} 	& - 	& 0.0010 	& \multirow6{*}{10} 	& \multirow6{*}{128} 	& \multirow6{*}{64} 	\\
		 & \textsc{Citeseer} 	& 0.05 	& 		 & - 	& 0.0010 	& 		 & 		 & 		 \\
		 & \textsc{PubMed} 	& 0.15 	& 		 & 64 	& - 	& 		 & 		 & 		 \\
		 & \textsc{Coauthor CS} 	& 0.10 	& 		 & 64 	& - 	& 		 & 		 & 		 \\
		 & \textsc{Amz Comp} 	& 0.05 	& 		 & - 	& 0.0010 	& 		 & 		 & 		 \\
		 & \textsc{Amz Photo} 	& 0.15 	& 		 & - 	& 0.0010 	& 		 & 		 & 		 \\
\hline\\[-0.7em]

\end{tabular}

}
\end{table}
\begin{table}[h]
    \centering
    \caption{Hyperparameters for DGI obtained by grid and random search.}
    \label{tab:hyper:dgi}
    \resizebox{
        \ifdim\width>\textwidth
            \textwidth
        \else
            \width
        \fi
    }{!}{
\begin{tabular}{llccccccc}
Diffusion & Dataset name & $\alpha$ & $t$ & $k$ & $\epsilon$ & \shortstack{Learning\\rate} & Encoder & \shortstack{Embedding\\dimension} \\
\hline\\[-0.7em]
\multirow6{*}{-} 	& \textsc{Cora} 	& \multirow6{*}{-} 	& \multirow6{*}{-} 	& \multirow6{*}{-} 	& \multirow6{*}{-} 	& \multirow6{*}{0.001} 	& \multirow6{*}{GCN} 	& \multirow6{*}{128} 	\\
		 & \textsc{Citeseer} 	& 		 & 		 & 		 & 		 & 		 & 		 & 		 \\
		 & \textsc{PubMed} 	& 		 & 		 & 		 & 		 & 		 & 		 & 		 \\
		 & \textsc{Coauthor CS} 	& 		 & 		 & 		 & 		 & 		 & 		 & 		 \\
		 & \textsc{Amz Comp} 	& 		 & 		 & 		 & 		 & 		 & 		 & 		 \\
		 & \textsc{Amz Photo} 	& 		 & 		 & 		 & 		 & 		 & 		 & 		 \\
\hline\\[-0.7em]
\multirow6{*}{Heat} 	& \textsc{Cora} 	& \multirow6{*}{-} 	& 1 	& - 	& 0.0001 	& \multirow6{*}{0.001} 	& \multirow6{*}{GCN} 	& \multirow6{*}{128} 	\\
		 & \textsc{Citeseer} 	& 		 & 5 	& - 	& 0.0001 	& 		 & 		 & 		 \\
		 & \textsc{PubMed} 	& 		 & 5 	& 64 	& - 	& 		 & 		 & 		 \\
		 & \textsc{Coauthor CS} 	& 		 & 5 	& 64 	& - 	& 		 & 		 & 		 \\
		 & \textsc{Amz Comp} 	& 		 & 1 	& - 	& 0.0001 	& 		 & 		 & 		 \\
		 & \textsc{Amz Photo} 	& 		 & 1 	& - 	& 0.0001 	& 		 & 		 & 		 \\
\hline\\[-0.7em]
\multirow6{*}{PPR} 	& \textsc{Cora} 	& 0.15 	& \multirow6{*}{-} 	& \multirow6{*}{-} 	& 0.0001 	& \multirow6{*}{0.001} 	& \multirow6{*}{GCN} 	& \multirow6{*}{128} 	\\
		 & \textsc{Citeseer} 	& 0.10 	& 		 & 		 & 0.0001 	& 		 & 		 & 		 \\
		 & \textsc{PubMed} 	& 0.15 	& 		 & 		 & 0.0010 	& 		 & 		 & 		 \\
		 & \textsc{Coauthor CS} 	& 0.15 	& 		 & 		 & 0.0010 	& 		 & 		 & 		 \\
		 & \textsc{Amz Comp} 	& 0.30 	& 		 & 		 & 0.0010 	& 		 & 		 & 		 \\
		 & \textsc{Amz Photo} 	& 0.30 	& 		 & 		 & 0.0010 	& 		 & 		 & 		 \\
\hline\\[-0.7em]

\end{tabular}

}
\end{table}

\end{document}